\documentclass[prb,twocolumn]{revtex4}
\usepackage{graphicx}
\usepackage{enumitem}
\usepackage{color}
\usepackage[colorlinks]{hyperref}
\usepackage{amsmath,amscd}
\usepackage{amssymb}
\usepackage[capitalise]{cleveref}
\pdfoutput=1

\begin{document}

\title{Nonlinear dynamics of Josephson vortices in a film screen under dc and ac magnetic fields}
\author{A. Sheikhzada}\email{ashei003@odu.edu}
\author{A. Gurevich}\email{gurevich@odu.edu}
\address{Department of Physics, Center for Accelerator Science, Old Dominion University,
Norfolk, VA 23529, USA}

\begin{abstract}
We present detailed numerical simulations of Josephson vortices in a long Josephson junction perpendicular to a thin film screen under strong dc and ac magnetic fields. By solving the sine-Gordon equation, we calculated the threshold magnetic field for penetration of fluxons as a function of frequency, and the power dissipated by oscillating fluxons as functions of the ac field amplitude and frequency. We considered the effects of superimposed ac and dc fields, and a bi-harmonic magnetic field resulting in a vortex ratchet dynamics. The results were used to  evaluate the contribution of weak-linked grain boundaries to the nonlinear surface resistance of polycrystalline superconductors under strong electromagnetic fields, particularly thin film screens and resonator cavities. 
\end{abstract}

\maketitle

\section{Introduction}
Dynamics of Josephson vortices in  long Josephson junctions (LJJs) under dc and ac magnetic fields has been the subject of much interest \cite{KL,BP,MT,LS,MS,FD,N}. For instance, the barrier and overlap LJJs have been studied extensively for applications in superconducting electronics \cite{KL,BP}, particularly flux flow oscillators \cite{ffo,KS,SS,SSb}. The electrodynamics of LJJ has attracted a renewed attention after the discovery of high-$T_c$ superconducting cuprates and iron based superconductors in which the grain boundaries between misoriented crystallites behave as long Josephson junctions which subdivide the materials into weakly coupled superconducting regions \cite{GB,DG}. The latter gives rise to the electromagnetic granularity \cite{HB} which is one of the serious obstacles for applications of the cuprate and the iron-based superconductors \cite{ag}.  

Another situation in which the weak-linked grain boundaries becomes essential occurs in superconducting resonator cavities \cite{cav} in which the amplitudes of the radio-frequency ($\simeq 0.1-5$ GHz) screening currents flowing at the inner surface of the cavity can approach the depairing current density $J_d$. In this case the grain boundaries even in such conventional materials as Nb can behave as LJJs \cite{gbi1,gbi2,gbi3,gbi4,gbi5}, even though they do not manifest themselves as weak links in dc magnetization or transport properties at much lower dc currents $J\ll J_d$.  It has been suggested that the Josephson vortices penetrating through grain boundaries can account for the linear decrease of the quality factors $Q(H_a)$ in Nb resonator cavities \cite{gbi4,gbi5,medQ}. Penetration of Josephson vortices under ac fields can also result in dissipation in polycrystalline thin films screen or multilayers \cite{ml}. Understanding the electrodynamics of Josephson vortices in weak-linked grain boundaries requires addressing the following issues: 1. The minimum amplitude of the ac field $H_p(\omega)$ the Josephson vortices start penetrating  the LJJ and the relation between $H_p$ and the dc lower critical field. 2. The field dependence of the power $P(H_a)$ dissipated in the LJJ at $H_a>H_p$ and its contribution to the nonlinear surface resistance $R_s(H_a)$. 3. The effect of a finite length of the LJJ on $H_p$ and $P(H_a)$ which would account for a finite grain size in polycrystalline materials of a finite film thickness in a screen \cite{ml}.  

In this paper we address a nonlinear electromagnetic response of a single LJJ across a thin film screen in a parallel field. We solved the sine-Gordon equation numerically to calculate the dynamics of penetration, annihilation and exit of Josephson vortices and antivortices oscillating under the ac field.  The paper is organized as follows. In Section \ref{sec:elj} we specify the main equations, the geometry and the boundary conditions. In Section \ref{sec:dcfield} we consider a LJJ in a thin film screen in a dc field and calculate the field-dependence of the power dissipated due to a net flow of Josephson vortices along the LJJ.   In Section \ref{sec:acfield} we consider a LJJ in a periodic ac field $H=H_a\sin\omega t$ and calculate the frequency dependence of the penetration field $H_p(\omega)$ for Josephson vortices, and the dissipated power as a function of $H_a$ and $\omega$. In the overdamped limit, the results of this section are in agreement with the previous works by McDonald and Clem\cite{Mc} and Zhai {\it et al.}\cite{Sr}, but are inconsistent with the assumption of Ref. \cite{gbi4,gbi5} that the surface resistance of a LJJ increases linearly with the amplitude of the rf field. In Section \ref{sec:asyacfield} we consider the response of the LJJ to an asymmetric ac magnetic field which causes a net force on the vortex, namely, a superposition of dc and ac fields, and a double-mode ac field which results in a dynamic ratchet effect. Section \ref{sec:diss} contains a discussion of the results.

\begin{figure}[h!]
\centering\includegraphics[width=8cm]{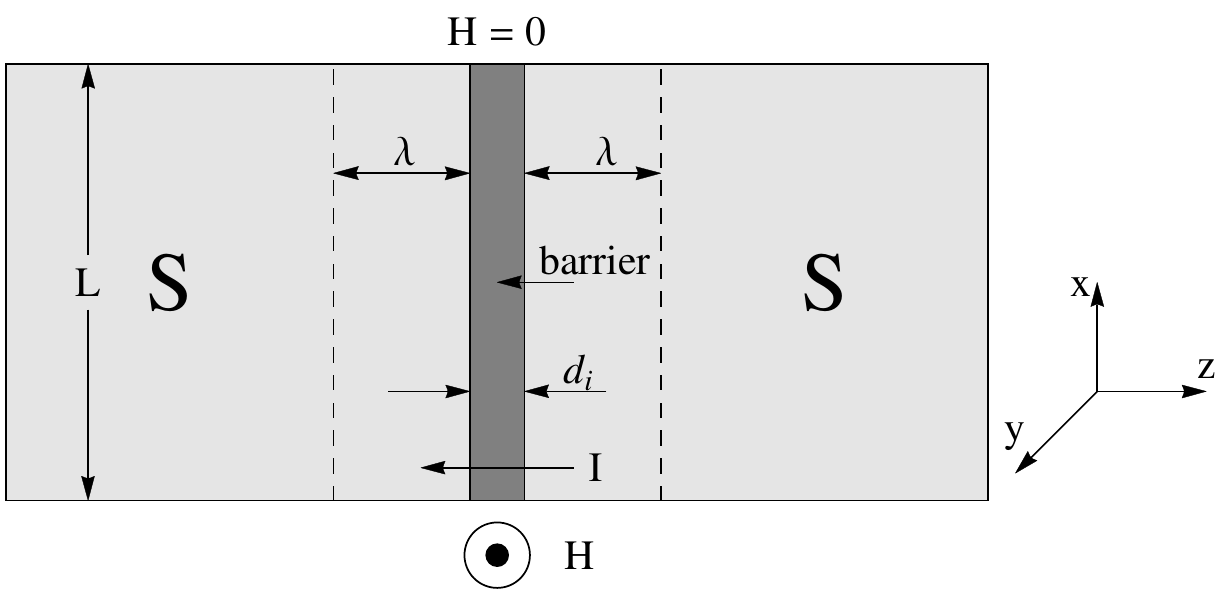}
\caption{Geometry of a long Josephson junction in a film which screens the uniform magnetic field $H$ applied in the region $x<0$.}
\label{fig1}
\end{figure} 

\section{Electrodynamics of a Long Junction}
\label{sec:elj}
We consider a LJJ perpendicular to a flat screen of width $L$ as shown in  \cref{fig1}. The uniform magnetic field $H(t)$ is applied along the $y$-axis parallel to one side of the screen 
at $x=0$. At the other side of the screen we assume the boundary condition $H(L,t)=0$. The LJJ is described by the sine-Gordon equation for the gauge-invariant phase difference 
$\gamma(x,t)$  \cite{KL,BP,MT} 
\begin{gather}
\lambda_J^{2}\gamma_{xx}=\sin\gamma+\omega_c^{-1}\gamma_t+\omega_p ^{-2}\gamma_{tt},
\label{eq} \\
\omega_p=\bigl(2\pi c J_c/\phi_0 C\bigr)^{1/2}, \qquad \omega_c=2\pi c R_i J_c/\phi_0.
\label{om}
\end{gather}
Here the subscripts $x$ and $t$ denote partial derivatives over $x$ and $t$, respectively, $\lambda_J=(c\phi_0/8\pi^2dJ_c)^{1/2}$ is the Josephson penetration depth, $J_c$ is the critical current density of the junction, $\omega_p$ is the Josephson plasma frequency, $\omega_c$ is the decrement due to quasiparticle ohmic currents, $c$ is the speed of light, $\phi_0$ is the magnetic flux quantum, $C$ is the specific capacitance of the junction, $R_i$ is the quasiparticle specific resistance per unit area, $d\approx 2\lambda$, and $\lambda$ is the London penetration depth.

As an illustration, we estimate $\omega_p$ and $\omega_c$ for Nb at different ratios of $J_c/J_d$ where $J_d=c\phi_0/12\sqrt{3}\pi^2\lambda^2\xi$ is the bulk depairing current density, and $\xi$ is the coherence length. Taking $\lambda\approx\xi\approx 40$ nm, the typical excess grain boundary resistance $R_i=2\times 10^{-13}$ $\Omega$m$^2$ for Nb \cite{gbn}, and $C=\epsilon/4\pi d_i$ where $\epsilon \simeq 3$ is the static dielectric constant of filled electron bands and $d_i\simeq 1$ nm is the atomic width of the grain boundary, we obtain $J_d\simeq 150$ MA/cm$^2$, and $R_iJ_d\simeq 0.3$V. Then  $\omega_c\simeq 10^{15}(J_c/J_d)$ Hz, and $\omega_p\simeq 4\cdot 10^{14} (J_c/J_d)^{1/2}$ Hz. The McCumber parameter 
$\beta_c=(\omega_c/\omega_p)^2\simeq 6J_c/J_d$ defines the effect of dissipation for a steady-state propagation of Josephson vortices; the case of $\beta_c\ll 1$ corresponds to the overdamped limit in which dissipative ohmic currents dominate over the displacement currents described by the inertial term $\propto \gamma_{tt}$ in  \cref{eq}.  
The Josephson weak link is by definition an interface with $J_c\ll J_d$ so, for the above numbers, the grain boundaries would be in the overdamped limit. However, $J_c$ across grain boundaries in Nb can be very high and close to $J_d\xi/\lambda$ in which case  \cref{eq} is no longer valid and the equation for $\gamma(x,t)$ becomes nonlocal  particularly in materials with large Ginzburg-Landau parameter $\lambda/\xi$ \cite{nje1,nje2,nje3}.   In this work we only consider the local Josephson limit described by  \cref{eq} both for $\beta_c>1$ and $\beta_c<1$.

For the geometry shown in  1, the local field distribution along the LJJ $B(x,t)=(\phi_0/4\pi\lambda)\gamma_x(x,t)$ defines the boundary conditions at $x=0$ and $x=L$: 
\begin{equation}
\gamma_x(0,t)=(4\pi\lambda/\phi_0)H(t),\qquad\gamma_x(L,t)=0.   
\label{bc}
\end{equation}

As will be shown below, penetration and annihilation of Josephson vortices can result in significant instant power dissipation $P(t)=\int_0^L V(x,t)J(x,t)dx$ 
per unit height of the junction along the $y$-axis, where $V=\phi_0\gamma_t/2\pi c$ is the voltage, and $J(x,t) $ is a sum of the Josephson, ohmic and displacement current densities:
\begin{equation}
P=\frac{\phi_0 J_c}{2\pi c}\int_{0}^{L} dx \left[\sin\gamma+\omega_c^{-1}\gamma_t+\omega_p^{-2}\gamma_{tt}\right]\gamma_t.
\label{P1}
\end{equation}
In a periodic ac field the contributions of Josephson and displacement currents vanish after  
averaging over the ac period $T=2\pi/\omega$. As a result, the average power is caused only by the ohmic currents:
\begin{equation}
\overline {P}=\frac{\phi_0 J_c}{2\pi c\omega_cT}\int_{t_0}^{t_0+T} dt \int_{0}^{L}\gamma_t^{2}dx.
\label{Pa}
\end{equation}

\section{Dc field}
\label{sec:dcfield}

The behavior of a LJJ in a static magnetic field is characterized by two field regions \cite{KL,BP}. At low fields $0<H<H_{c1J}$ the LJJ is in a Meissner state in which the local magnetic field $B(x) = \phi_0\gamma_x/4\pi\lambda$ is screened at the edge of the junction over the length $\sim\lambda_J$. At high fields, $H>H_{c1J}$ penetration of Josephson vortices each carrying the flux quantum $\phi_0$ becomes thermodynamically favorable. There is also the field region $H_{c1J}<H<H_1$ of metastable Meissner state, where $H_1=\pi H_{c1J}/2$ plays the role of a superheating field at which the edge energy barrier for the penetration of Josephson vortices disappears.  Here $H_{c1J}$ and $H_1$ are given by
\begin{equation}
H_{c1J}=\frac{\phi_0}{\pi^2\lambda\lambda_J},\qquad H_1=\frac{\phi_0}{2\pi\lambda\lambda_J}.
\label{hc1}
\end{equation} 

To calculate the dynamics of penetration of vortices at $H>H_1$ we solve  \cref{eq} numerically.
It is convenient to write  \cref{eq} in a dimensionless form, using the rescaled variables $x\to x/\lambda_J$ and $t\to\omega_p t$: 
\begin{equation}
\gamma_{tt}+\alpha\gamma_t=\gamma_{xx}-\sin\gamma,
\label{eqd}
\end{equation}
where $\alpha=1/\sqrt{\beta_c}=\omega_p/\omega_c$. The boundary conditions \cref{bc} and the instant power \cref{P1} take the form
\begin{eqnarray}
\gamma_x(0,t)=h,\qquad\gamma_x(l,t)=0,
\label{bcd}    \\
P/P_0=\int_{0}^{l} dx \left[\sin\gamma+\alpha\gamma_t+\gamma_{tt}\right]\gamma_t,
\end{eqnarray}
where $l=L/\lambda_J$, $h=(4\pi\lambda \lambda_J/\phi_0)H$, $P_0={\overline{c}\phi_0 J_c }/{2\pi c}$ and 
$\overline{c}\equiv\omega_p\lambda_J=c/(8\pi\lambda C)^{1/2}$ is the Swihart velocity \cite{BP,MT}.

\begin{figure}[h!]
\centering\includegraphics[width=8cm]{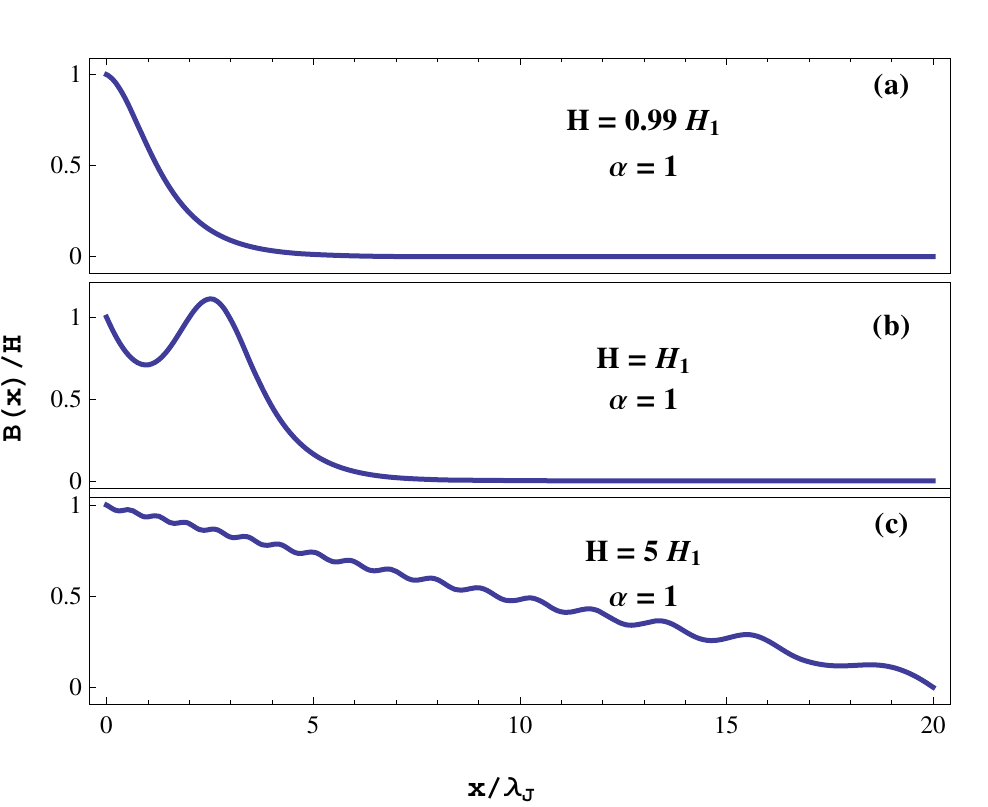}
\caption{(on the Web only) Profiles of the local magnetic field $B(x)$ in a LJJ at $L=20\lambda_J$, $\alpha=1$ and different values of $H$.}
\label{fig2}
\end{figure}

We first consider the overdamped limit in which the term $\propto \gamma_{tt}$ can be neglected and  \cref{eqd} turns into a nonlinear diffusion equation,
\begin{equation}
\alpha\gamma_t=\gamma_{xx}-\sin\gamma.
\label{eqdo}
\end{equation}
 Solutions of  \cref{eqdo} are shown in  \cref{fig2} for $\alpha= 1$, $L=20\lambda_J$ and different magnetic fields $H$.   \cref{fig2}(a) shows the metastable Meissner state at $H_{c1J}<H<H_1$, in which the magnetic flux is screened at the edge over the length $\sim \lambda_J$.  \cref{fig2}(b) shows the case of $H = H_1$ at which the first vortex nucleates at the edge and accelerates until the velocity becomes limited by the friction force of ohmic currents.  \cref{fig2}(c) shows the flux flow state at higher field $H=5H_1$ at which vortices periodically enter, travel all the way to the other end at $x=L$ where they disappear. This dynamic state  is characteristic of a thin film screen \cite{ml} in which the LJJ provides a path for constant flux pumping from the region of applied field to the inner region of $H=0$, unlike a stationary chain of Josephson vortices in a uniform field which is the same of both edges of the junction \cite{KL,BP}.    

\begin{figure}[h!]
\centering\includegraphics[width=7.5cm]{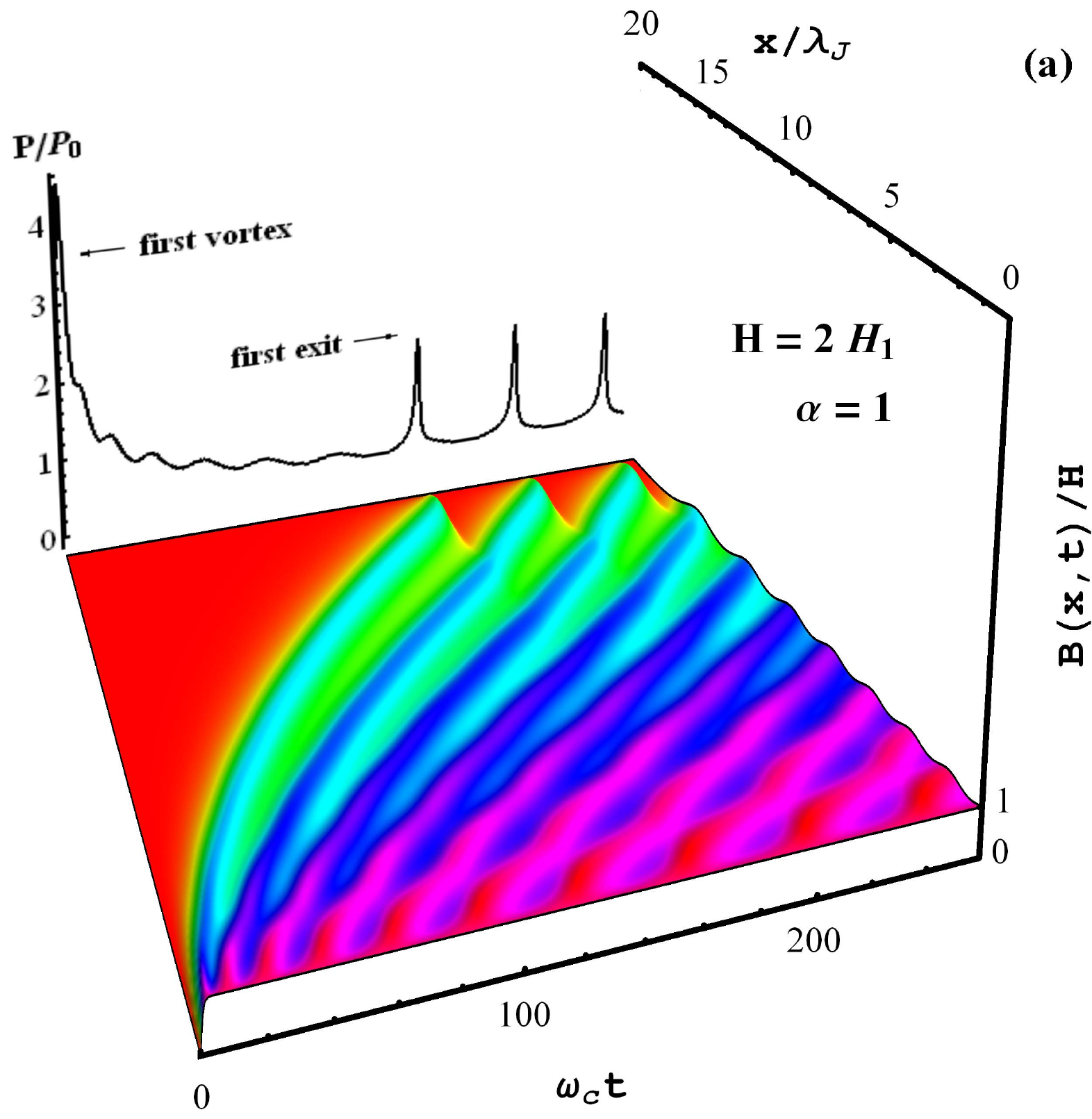}
\bigskip
\centering\includegraphics[width=7.5cm]{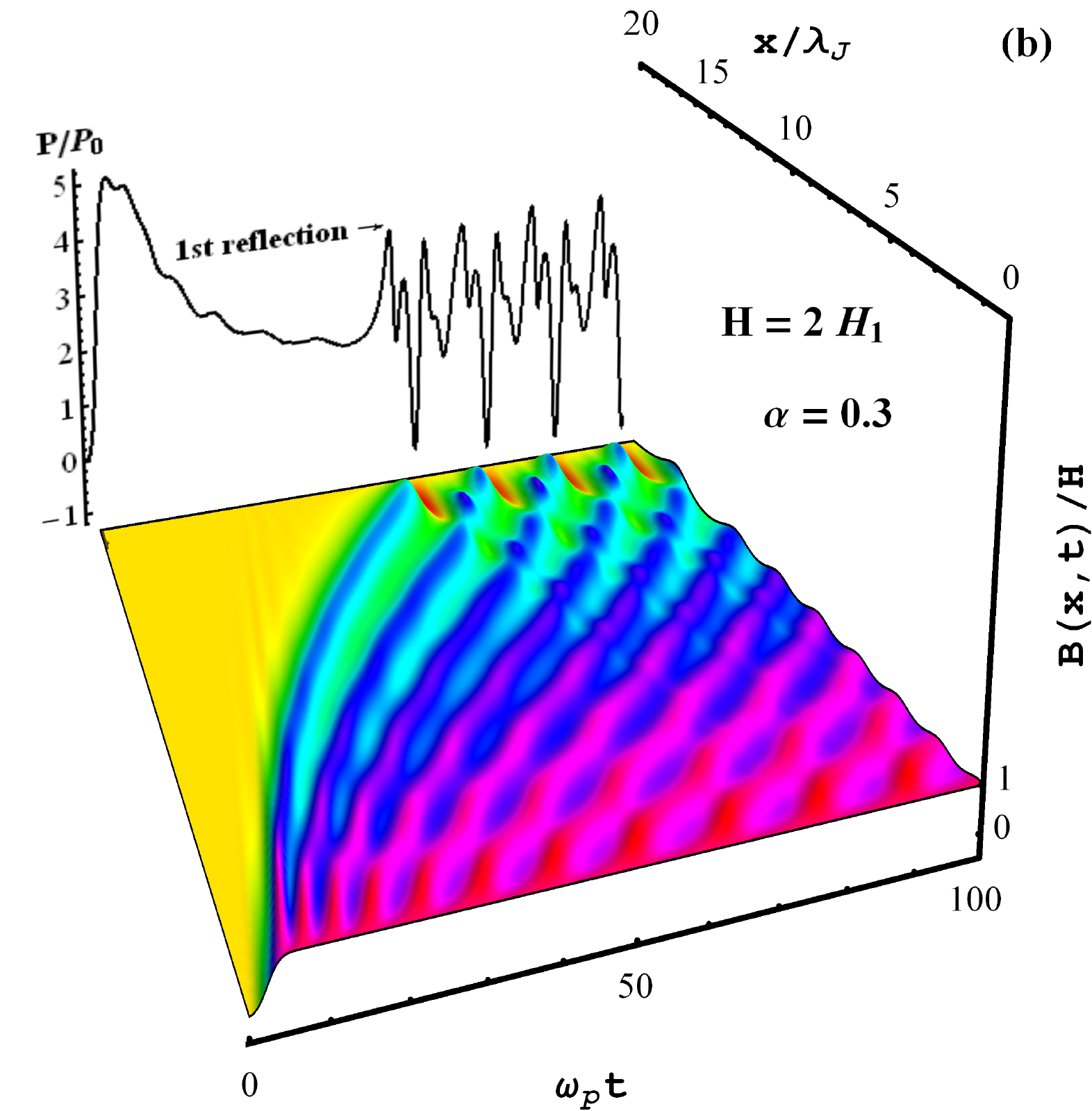}
\caption {(on the Web only) Evolution of the local magnetic field $B(x,t)$ along the LJJ, and the instant power $P(t)$ calculated for $H=2H_1$:  (a) Results of solution of  \cref{eqdo} in the 
overdamped limit at $\alpha=1$; (b) Results of solution of  \cref{eqd} for a moderately dissipative case of $\alpha=0.3$. Standing electromagnetic waves generated by moving 
vortices in the LJJ manifest themselves in "ripple" on $B(x,t)$ and in a more complex behavior of $P(t)$ than for the overdamped limit.  }
\label{fig3}
\end{figure}

 \cref{fig3}(a) shows the evolution of the local field $B(x,t)$ along the LJJ and the instant power $P(t)$ at $H=2H_1$ and $\alpha=1$. One can see that each penetration and annihilation of vortices at the edges produces peaks in $P(t)$. The highest peak in $P(t)$ occurs during penetration of the first vortex at $x=0$ after the field was turned on and the vortex is accelerated strongly by the Lorentz force of screening current. After penetration of several vortices, the Lorentz force which pushes the next vortex in the junction is reduced by the counterflow of vortices already in the LJJ, so the peaks in $P(t)$ caused by penetrating vortices are reduced.  As the vortex exits the junction at $x=L$, it is accelerated again due to attraction to its antivortex image  \cite{MS}, producing peaks in $P(t)$.  

\begin{figure}[h!]
\centering\includegraphics[width=7.5cm]{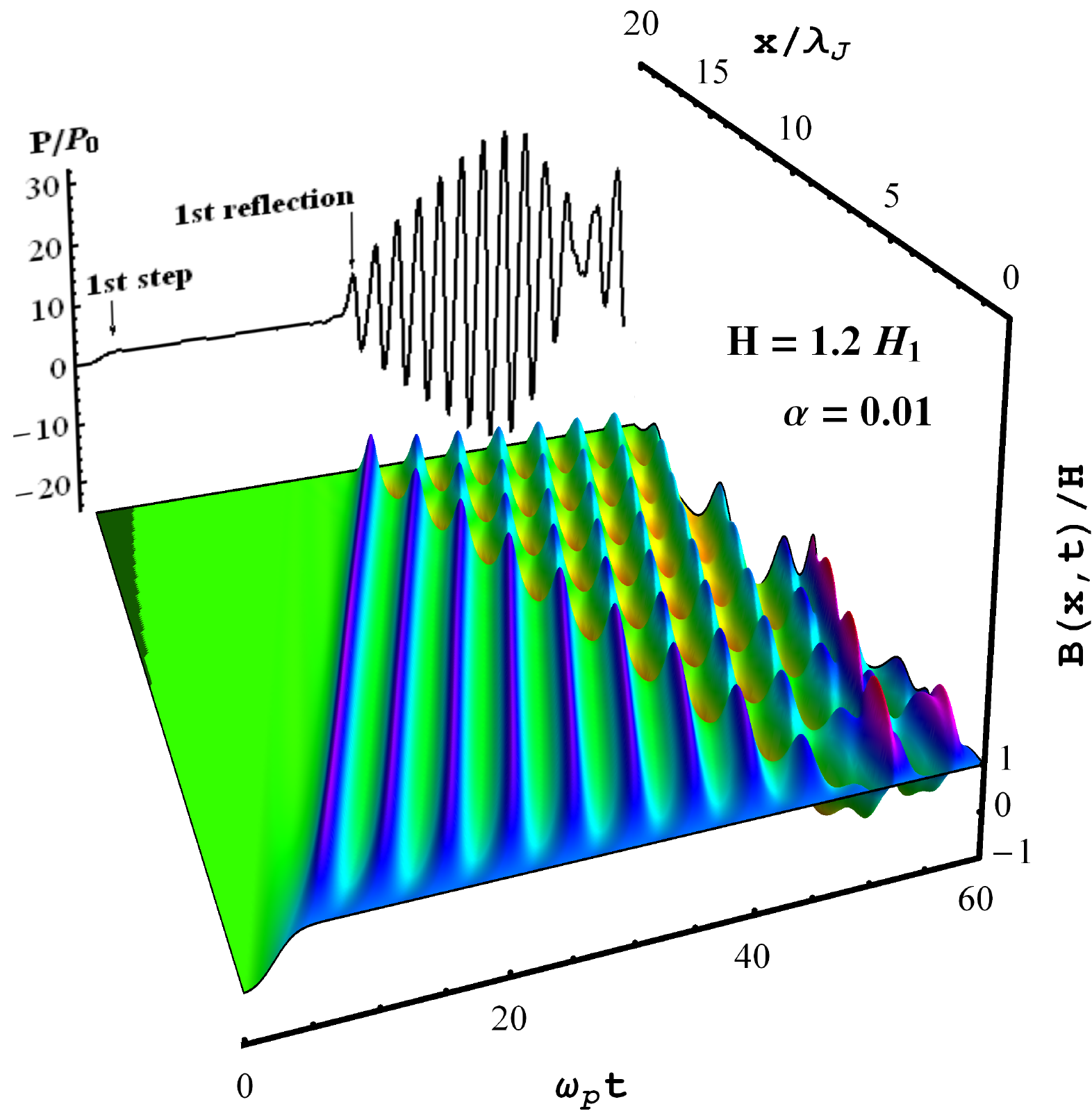}
\caption {(on the Web only) Evolution of the local magnetic field $B(x,t)$ calculated for a weakly dissipative case of $\alpha=0.01$ at $H=1.2H_1$. Vortices undergo multiple reflections from the edges with the reversal of their polarity and velocity.}
\label{fig4}
\end{figure}

Now we consider the effect of displacement currents on dynamics of Josephson vortices by first solving the full  \cref{eqd} for moderate damping at $\alpha=0.3$ and $H=2H_1$. The results shown in  \cref{fig3}(b) indicate that in this case vortices gain some inertia and upon reaching the edges dissipate most of their energy, but a small part of it would get reflected in the form of decaying electromagnetic waves back to the junction. For weaker damping $(\alpha<0.1)$, vortices move with a nearly uniform velocity until they get reflected from the edge of the junction without losing much of their energy but reversing their polarity and velocity\cite{FD}. As shown in  \cref{fig4}, for $\alpha=0.01$ and $H=1.2H_1$, vortices move almost with their initial velocity but upon reaching the edge of the junction at $x=L$, they get reflected as anti-vortices. The reflected anti-vortices pass through incoming vortices\cite{LS} causing only small amount of dissipation. The multiple reflections of vortices from the edges along with continuous pumping of the electromagnetic waves can result in a chaotic behavior of $\gamma(x,t)$ which we do not address in this work.

\begin{figure}[h!]
\centering\includegraphics[width=8cm]{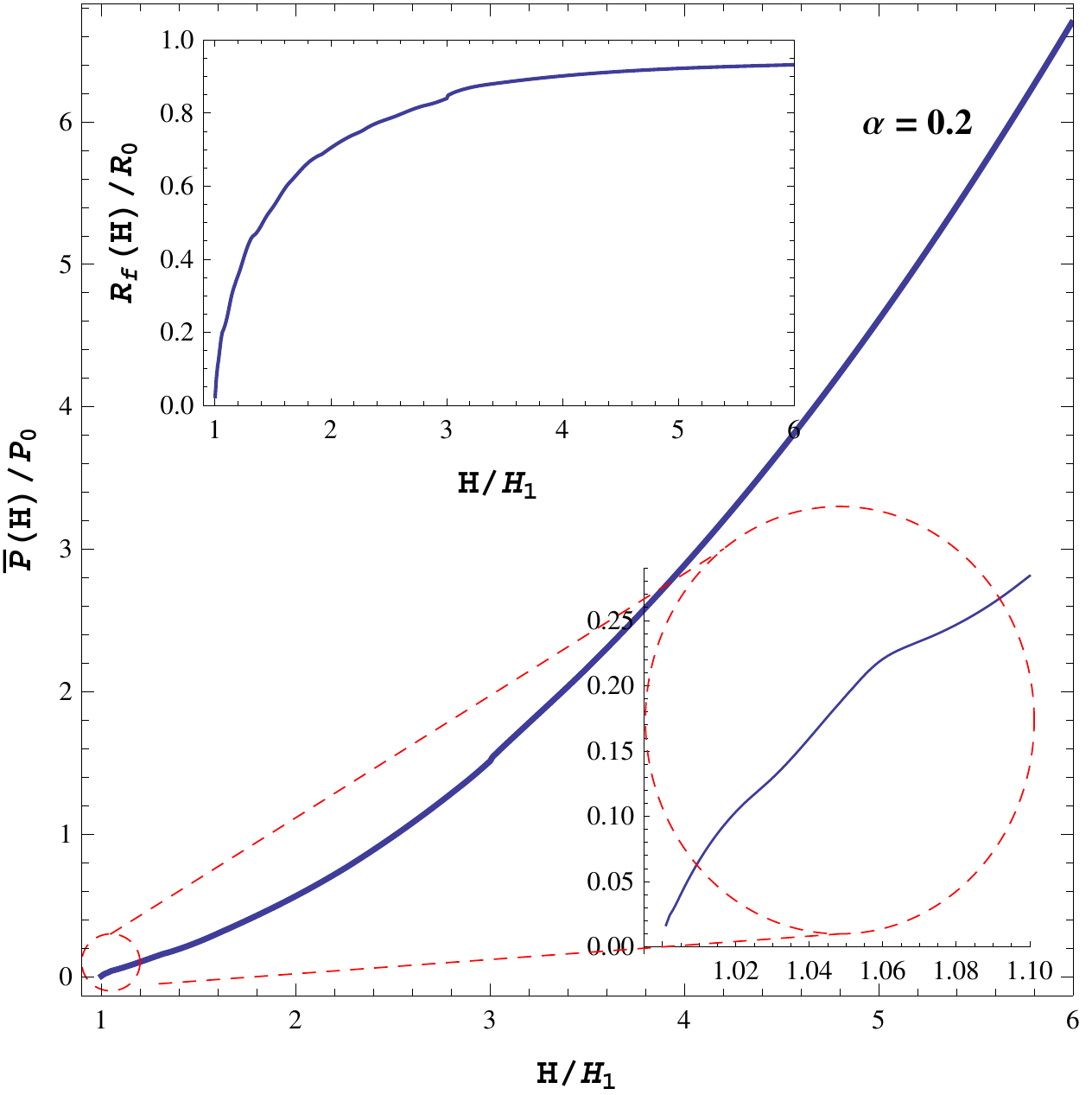}
\caption {(on the Web only) Averaged power $\bar{P}/P_0$ as a function of reduced dc magnetic field $H/H_1$ calculated for $\alpha=0.2$. Inset shows the flux flow resistance as a function of $H/H_1$ where $R_0=R_i/L$ is the total quasiparticle resistance of the junction. }
\label{fig5}
\end{figure}

 Shown in  \cref{fig5} is the averaged power $\bar{P}$ generated by moving Josephson vortices calculated from  \cref{eqd} at $\alpha=0.2$. At high fields $[H> (3-4)H_1]$, the dependence $\bar{P}(H)$ becomes nearly quadratic in $H$ but at lower fields,  there are step-like features in $\bar{P}(H)$ associated with penetration of Josephson vortices. From the power dissipation relation $\overline{P}=I^2R_f$ where $I=cH/4\pi$ is the total current flowing along the screen, we calculate the field dependence of the flux flow resistance $R_f(H)$ plotted in inset of  \cref{fig5}. Here $R_f(H)$ vanishes at $H=H_1$ and increases with $H$, approaching the total LJJ resistance $R_0=R_i/L$ at $H>4H_1$.

\section{Single-mode ac field}
\label{sec:acfield}

In this section we consider a LJJ under a single-mode ac magnetic field, $H=H_a \sin{\omega t}$. In this case it is more convenient 
to rescale the time in the units of the ac period $t\rightarrow\omega t$, so that the dimensionless sine-Gordon equation takes the form
\begin{equation}
\beta\gamma_{tt}+\alpha\gamma_t=\gamma_{xx}-\sin\gamma,
\label{eqa}
\end{equation}
where $\alpha=\omega/\omega_c$ and $\beta=(\omega/\omega_p)^2$. The boundary conditions become
\begin{equation}
\gamma_x(0,t)=h_a\sin t,\quad\gamma_x(l,t)=0,   
\label{bca} 
\end{equation}
where  $h_a=(4\pi\lambda\lambda_J/\phi_0)H_a$. The instant power is then
\begin{equation}
P/P_0=\int_{0}^{l} dx \left[\sin\gamma+\alpha\gamma_t+\beta\gamma_{tt}\right]\gamma_t,
\label{pa}
\end{equation}
where $P_0=\phi_0 J_c \lambda_J \omega/2\pi c$.

As was shown above, the plasma frequency for the grain boundaries in Nb is typically in the infrared region ($\omega_p\sim 10^{12}-10^{14}$ Hz) so for many microwave and rf applications ($\omega\sim 0.1-10 $ GHz), the parameter $\beta\ll \alpha$(i.e. $\omega\omega_c\ll \omega_p^2$) is negligible and  \cref{eqa} reduces to 
\begin{equation}
\alpha\gamma_t=\gamma_{xx}-\sin\gamma,
\label{eqao}
\end{equation}
 
\begin{figure}[h!]
\centering\includegraphics[width=8cm]{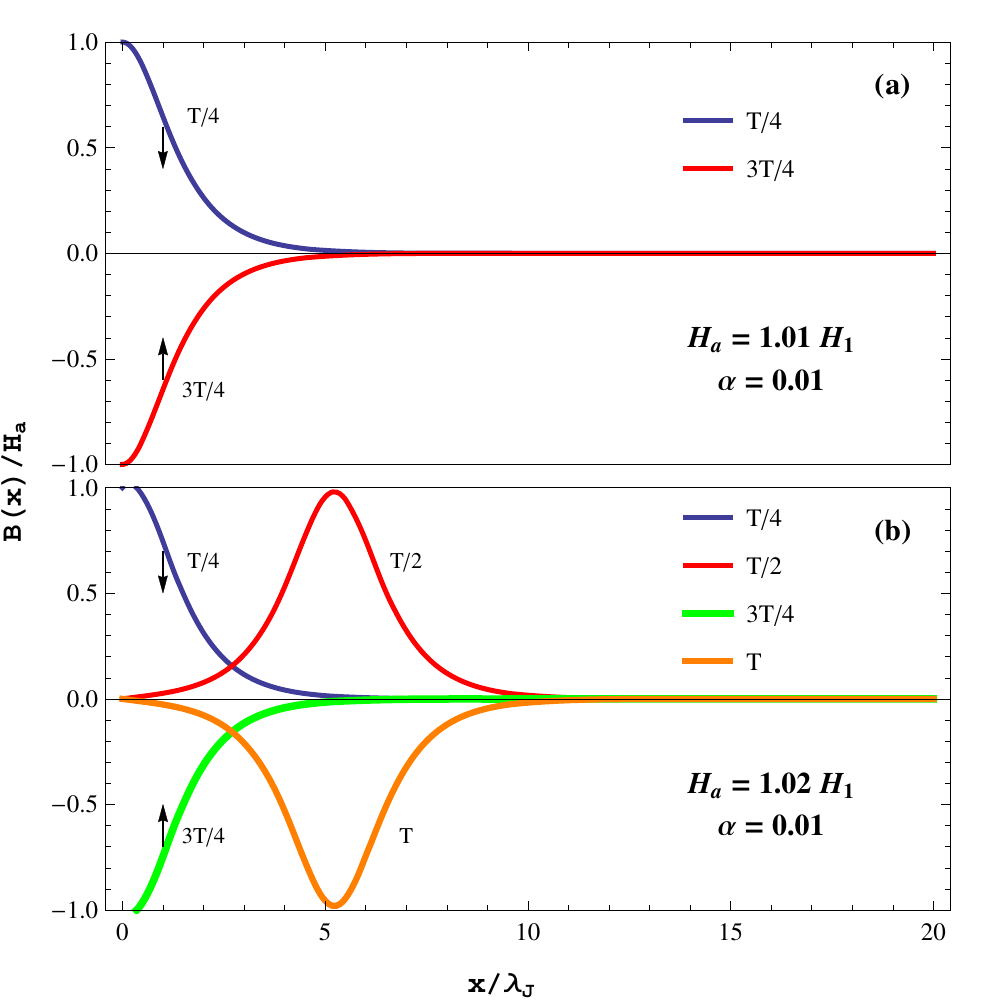}
\caption{(on the Web only) Snapshots of magnetic field profiles $B(x)$ calculated for $\alpha=0.01$ at different times: (a) just before the first vortex penetrates; (b) just after the penetration of the first vortex/anti-vortex occurred. Arrows show whether the applied field $H(t)$ is increasing or decreasing.}
\label{fig6}
\end{figure}

Our numerical simulations of  \cref{eqao} have shown that it has a solution $\gamma(x,t)$ with the periodicity of the applied ac field. Shown in  \cref{fig6} are the profiles of magnetic field just before and after penetration of a vortex calculated for $\alpha=0.01$. These snapshots of $B(x,t)$ at different times and $H_a\approx H_1$ suggest that a vortex (or antivortex during the negative field cycle) get trapped at the edge of the junction, just because vortices under oscillating ac field have limited time to enter the junction.  As a result, the threshold field $H_p(\omega)$ of vortex penetration becomes larger than $H_1$ and increases with the frequency, so that there is enough time during the part of the period when $H_a|\sin\omega t| > H_1$ for the vortex to penetrate by the distance $\sim\lambda_J$. Calculations of $\bar{P}$ given below show that the rf power dissipated in the LJJ increases sharply at $H_a>H_p$. 

\subsection{Flux dynamics}

\begin{figure}[h!]
\centering\includegraphics[width=7.5cm]{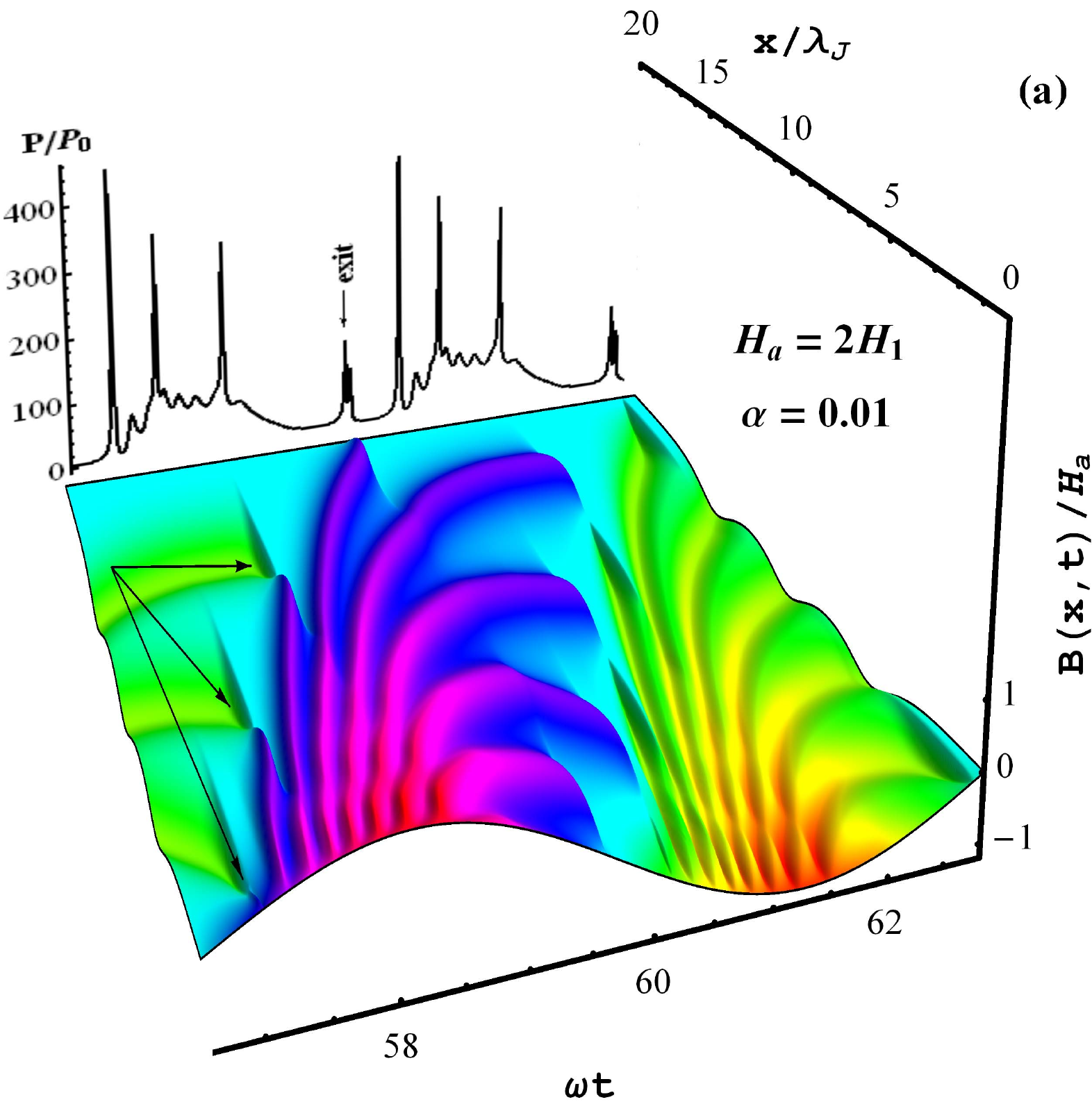}
\bigskip
\centering\includegraphics[width=7.5cm]{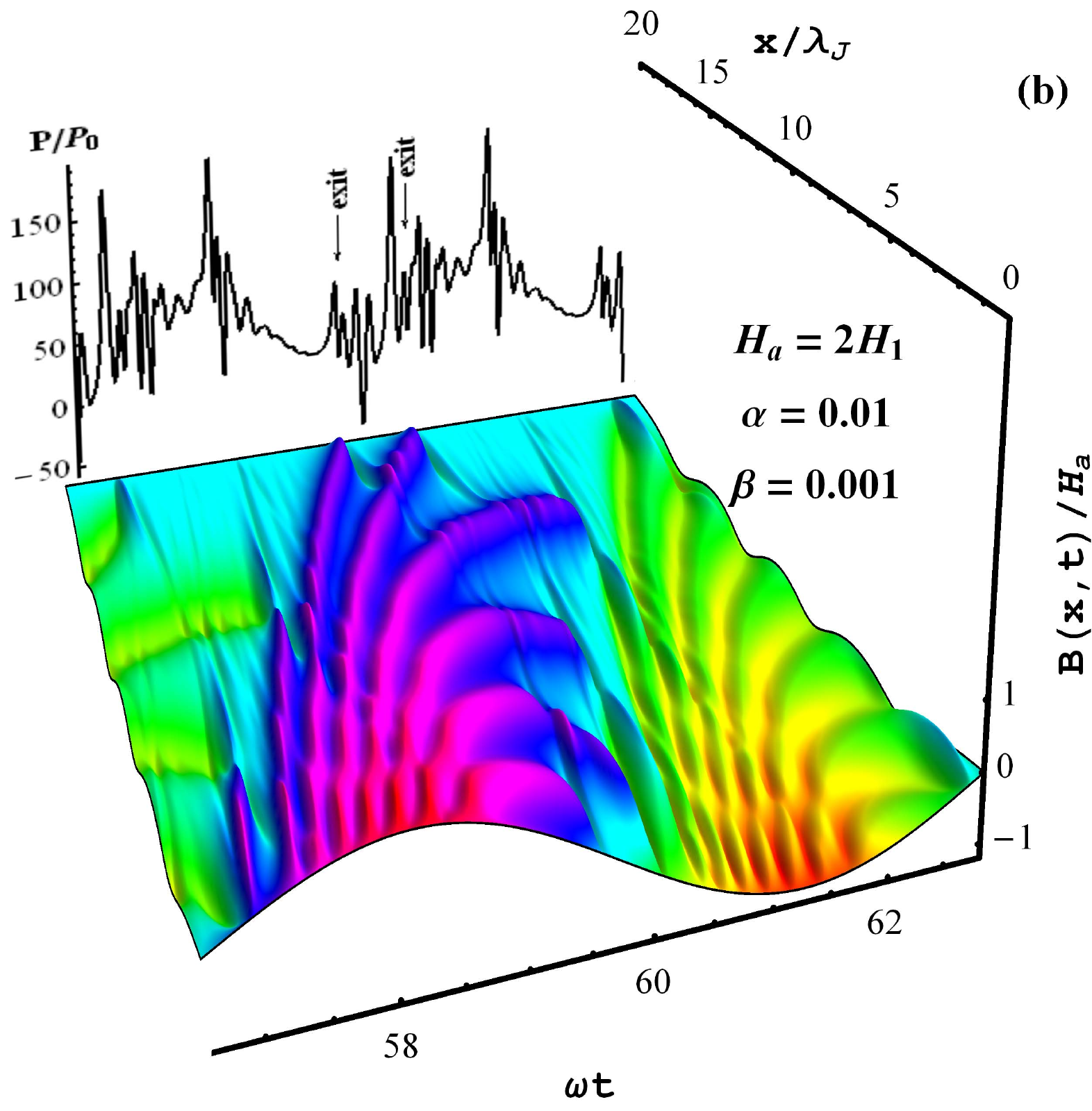}
\caption {(on the Web only) Evolution of the local magnetic field $B(x,t)$ and the instant dissipated power calculated from  \cref{eqa} for $H_a=2H_1$:  (a) Overdamped limit at $\alpha=0.01$. Arrows show the points of annihilation of vortices and antivortices; (b) moredrately overdamped limit at $\alpha=0.01$ and $\beta=0.001$. The flux dynamics is similar to (a) except the additional "ripples" on $B(x,t)$ due to electromagnetic waves generated because of the effect of vortex inertia. }
\label{fig7}
\end{figure}

Unlike the unidirectional flow of vortices under dc field considered in section \ref{sec:dcfield}, the flux dynamics under ac field 
includes penetration of Josephson vortices during the positive ac cycle followed by penetration of antivortices during the negative cycle and 
their subsequent annihilation.  Shown in  \cref{fig7} are representative examples of the evolution of the local 
 magnetic field $B(x,t)$, and the corresponding instant power $P(t)$ plotted for a full ac cycle calculated from  \cref{eqa} at $H_a=2H_1$. 
 In the particular case of overdamped flux dynamics shown in  \cref{fig7}(a),  about eight vortices penetrate the LJJ during the positive ac cycle, giving rise to small peaks in $P(t)$. Of these  vortices, the first three annihilate upon collisions with residual anti-vortices generated during the previous negative half cycle, while the fourth one goes all the way along the junction and exits at the other end. The last four vortices do not reach the end of the junction and turn around as $H(t)$ changes sign; the very last vortex exits before antivortices appear, but three other vortices annihilate on their way back with incoming anti-vortices generated during the negative ac cycle. The same process repeats for antivortices during the negative cycle. Notice that annihilation of vortices and antivortices inside the junction results in peaks in $P(t)$ that are significantly higher than the peaks in $P(t)$ during penetration or exit of vortices at the edges.

\cref{fig7}(b) shows the effect of displacement currents on the flux dynamics in a moderately overdamped limit at $\beta\ll\alpha$. We found that if $\beta\lesssim 0.1\alpha$, the response of the junction to the ac field remains periodic and similar to the solutions at $\beta\to 0$, except for generation of electromagnetic waves by accelerating/decelerating vortices upon interaction with boundaries and other vortices. As shown in  \cref{fig7}(b), the number of vortices does not change as compared to  \cref{fig7}(a), but because they now have some inertia, two vortices are able to reach the edge and leave behind weak electromagnetic radiation which manifests itself in "ripple" on $B(x,t)$ and a more irregular behavior of $P(t)$. 

\begin{figure}[h!]
\centering\includegraphics[width=7.5cm]{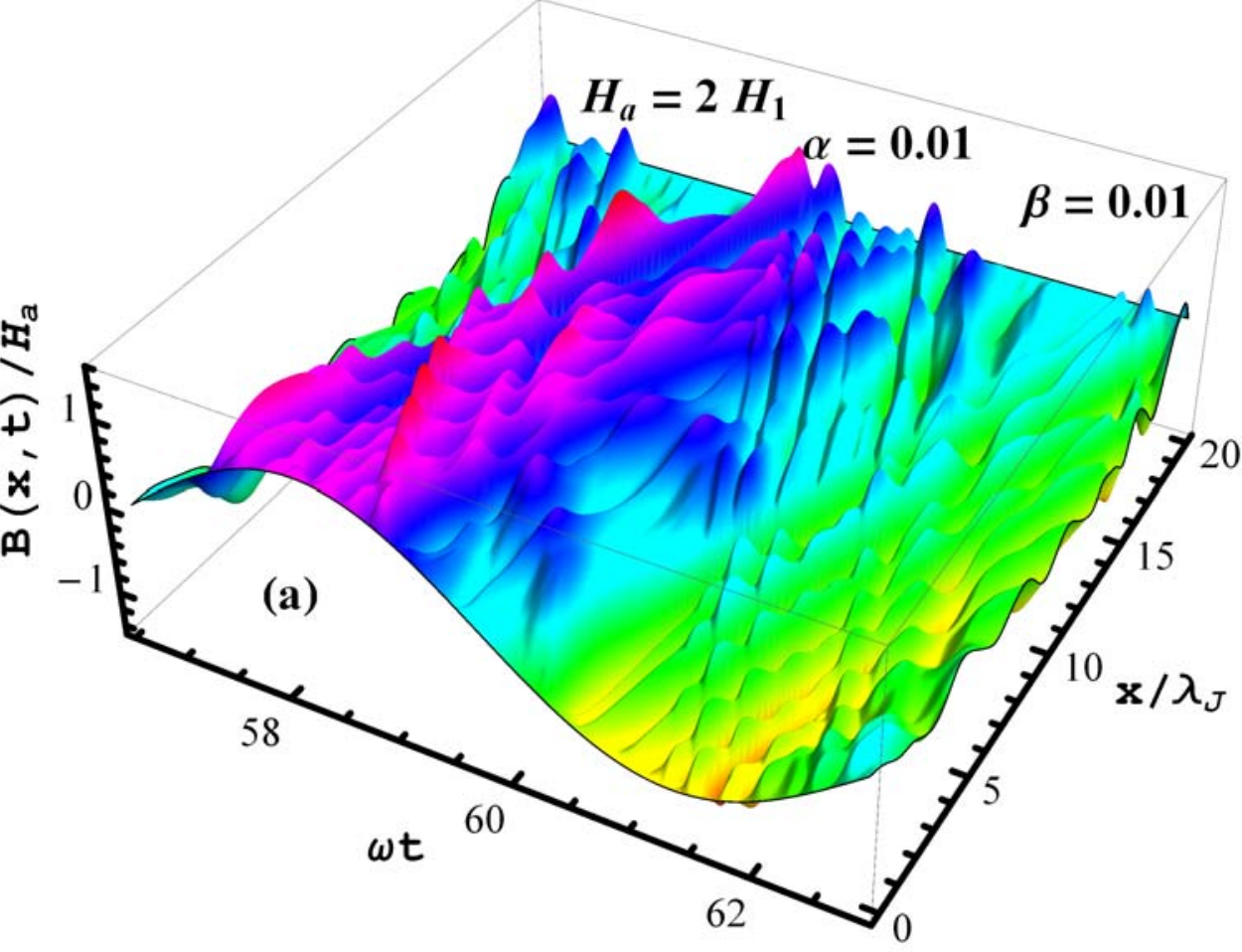}
\bigskip
\centering\includegraphics[width=7.5cm]{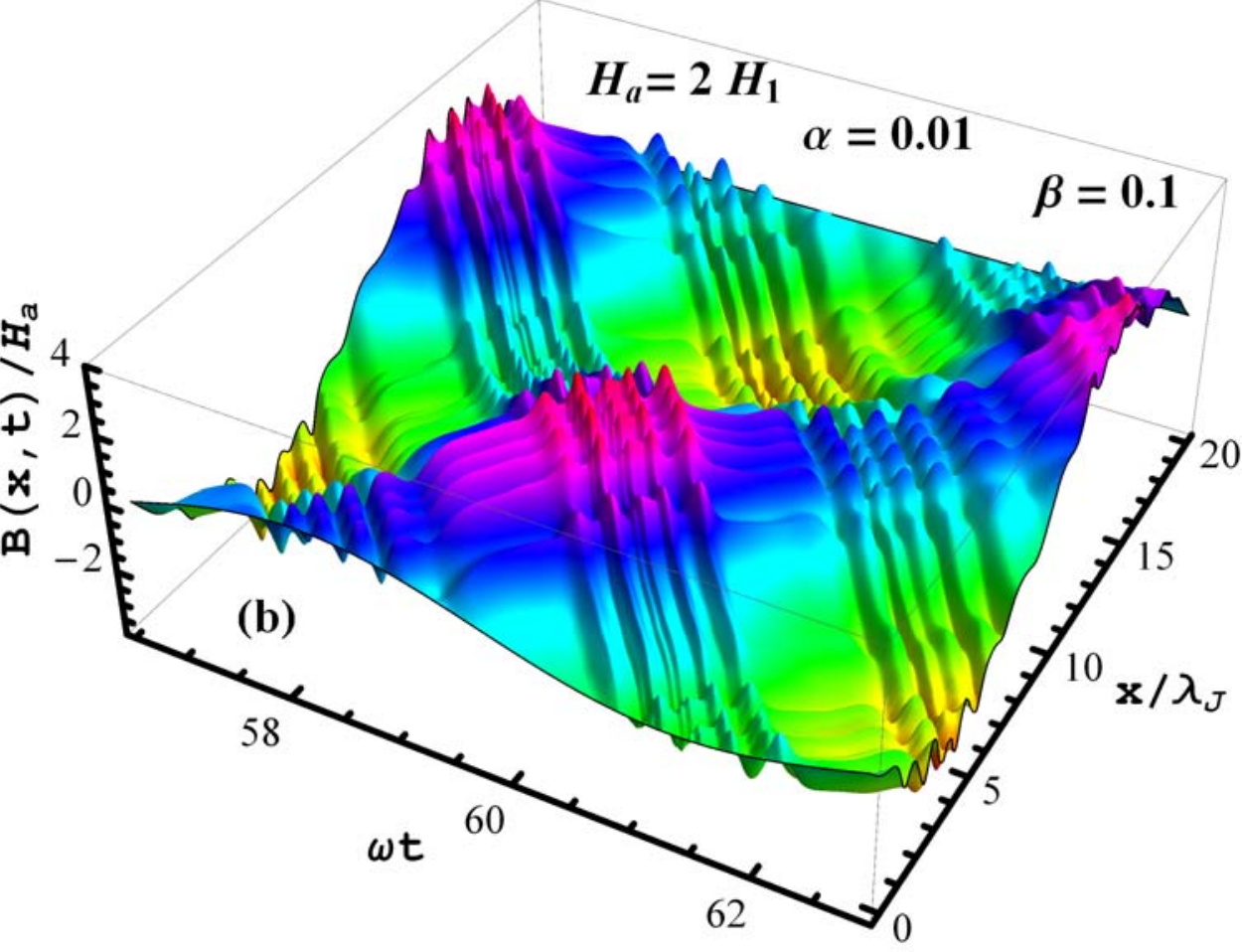}
\caption {(on the Web only) Evolution of the local magnetic field $B(x,t)$ at $H_a=2H_1$, and $\alpha=0.01$ for different values of $\beta$: (a) $\beta=0.01$; (b): $\beta=0.1$. In both cases the 
ripple on $B(x,t)$ is due to standing electromagnetic waves generated by accelerating/decelerating vortices. In a weakly dissipative case shown in (b), vortex/anti-vortex bundles 
form during each half cycle. }
\label{fig8}
\end{figure}

In the case of $\beta\sim\alpha$ shown in  \cref{fig8} vortices quickly enter the junction during the positive ac cycle and move with a nearly uniform velocity until they hit  the other edge. There they get reflected from the edge as anti-vortices which then collide with newly entered anti-vortices,  giving rise to local spikes of high magnetic field inside the junction before passing through each other and making their trip toward the other edge. In this regime, vortex dynamics is getting more chaotic as depicted in  \cref{fig8}(a); after several reflections, vortices eventually lose their energy due to ohmic losses and exit. \cref{fig8}(b) illustrates a more chaotic behavior at $\beta>\alpha$; here vortices undergo more reflections and less dissipation, forming  a dynamic pattern in which twice in every ac period, half of the junction is filled with vortices and half with anti-vortices. 

\subsection{Dissipated power}

\begin{figure}[h!]
\centering\includegraphics[width=8cm]{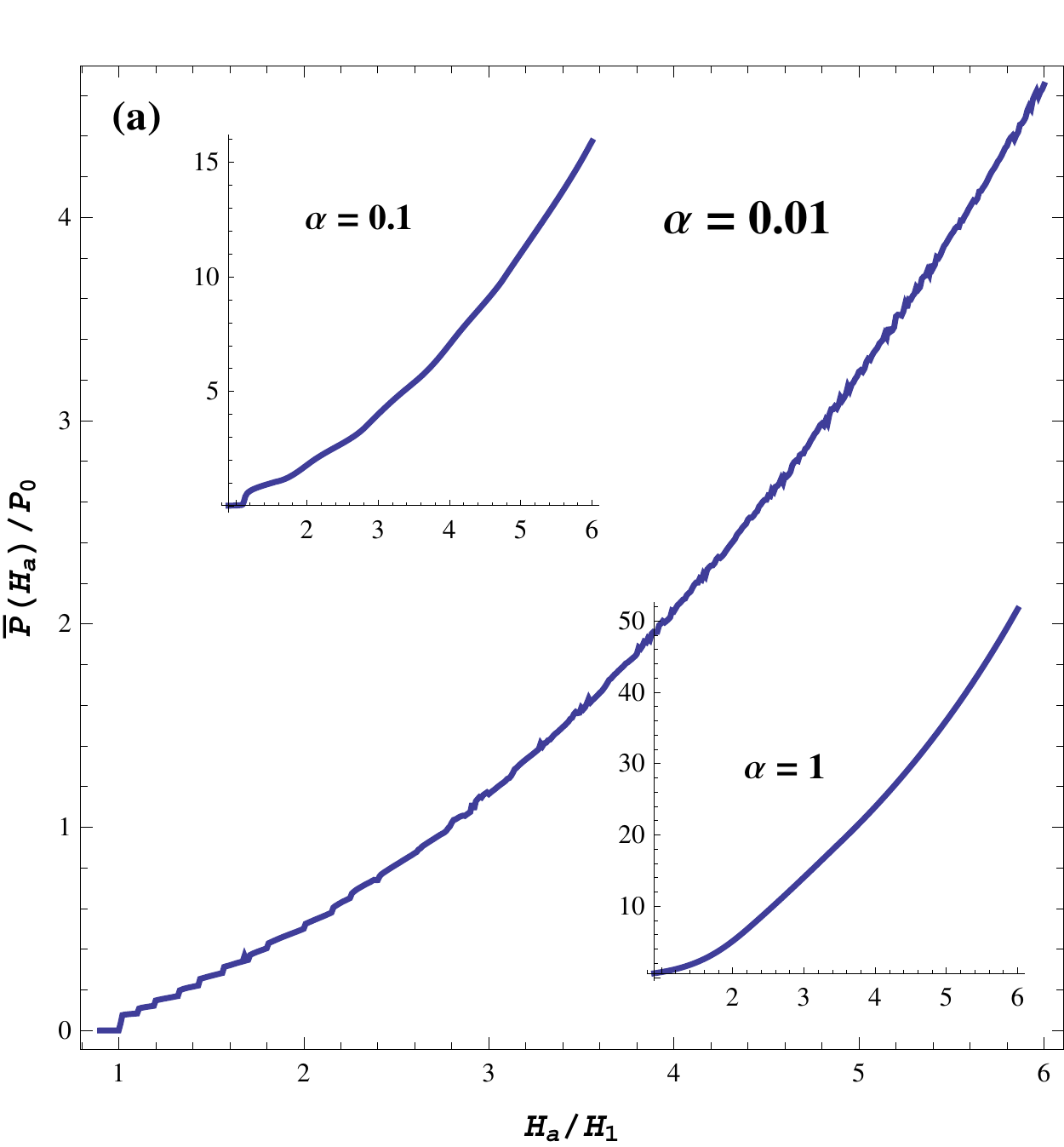}
\bigskip
\centering\includegraphics[width=8cm]{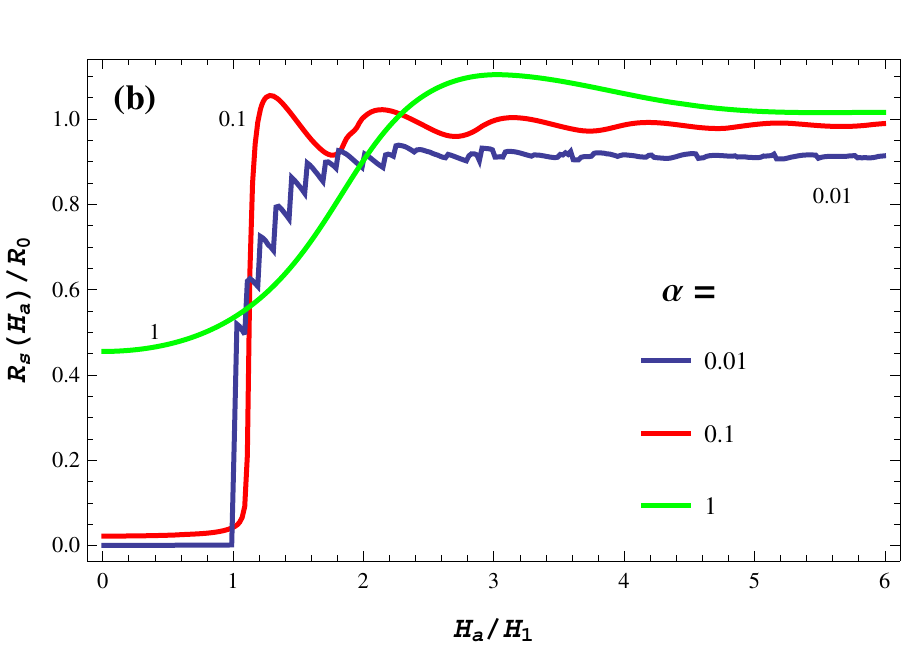}
\caption {(on the Web only) (a) Plots of $\bar{P}(H_a)$ for different dimensionless frequencies $\alpha=\omega/\omega_c=0.01, 0.1$ and $1$, in the overdamped limit. Steps in $\bar{P}(H_a)$ are associated with the addition of vortices to the junction. (b) The surface resistance  $R_s(H_a)$ for different $\alpha$.  }
\label{fig9}
\end{figure}

We now calculate the mean dissipated power $\bar{P}$ in the overdamped limit ($\beta\ll\alpha$), by averaging  \cref{pa} over the ac period: 
\begin{equation}
\overline{P}/P_0=\frac{\alpha^2}{2\pi}\int_{0}^{2\pi} dt \int_{0}^{l}\gamma_t^2 dx,
\end{equation}
where $P_0= \phi_0 J_c \lambda_J \omega_c/2\pi c$. Plotted in  \cref{fig9}(a) is $\bar{P}$ as a function of ac field amplitude for different values of the dimensionless frequency $\alpha=\omega/\omega_c$ in the overdamped limit. One can clearly see steps in $\bar{P}(H_a)$ due to the change of the mean number of vortices in the junction as $H_a$ increases. For smaller frequencies $\alpha$, the steps are sharper and decrease in amplitude as $H_a$ increases. As $\alpha$ increases, the sharp steps become broader until they disappear completely and $\bar{P}(H_a)$ quickly becomes quadratic in $H_a$. 

\begin{figure}[h!]
\centering\includegraphics[width=8cm]{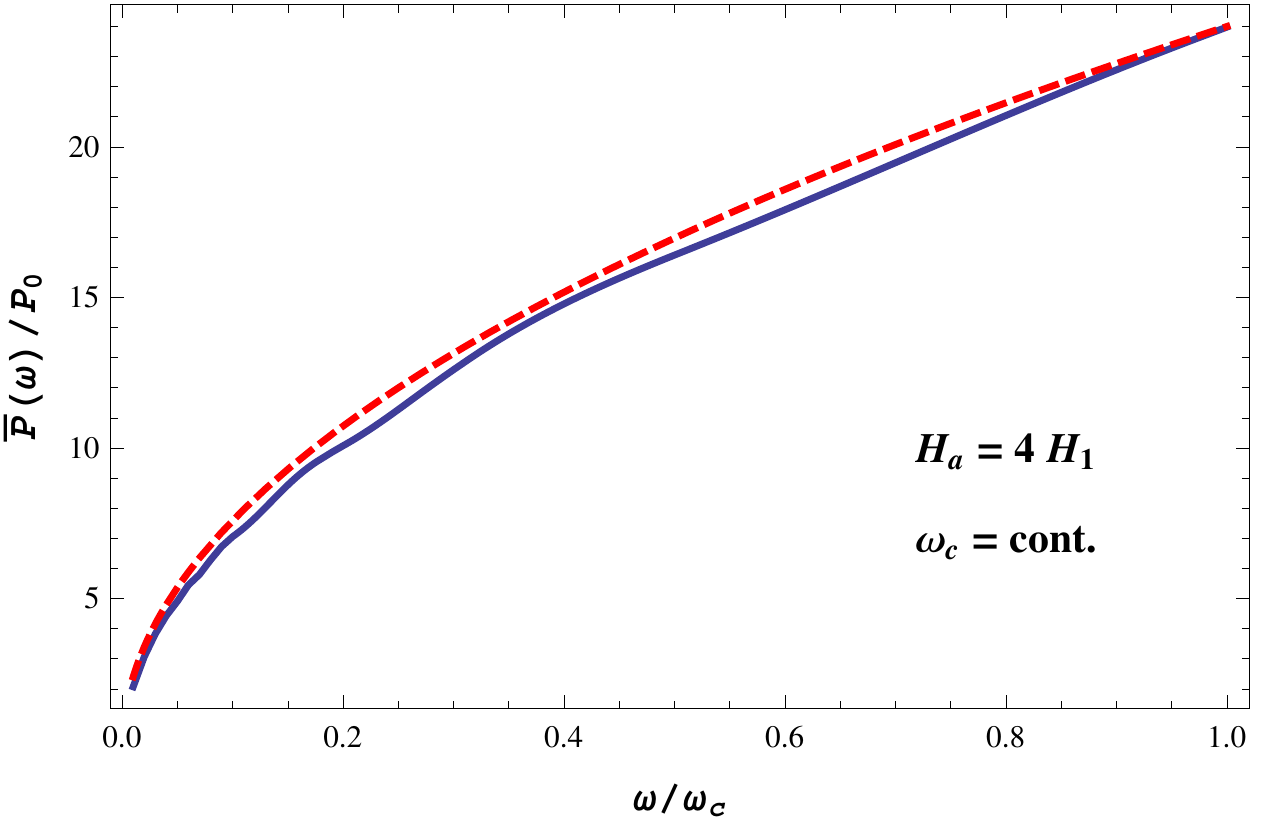}
\caption {(on the Web only) Plot of $\bar{P}(\omega)$ at $H_a=4 H_1$. The dashed line shows a square root function. }
\label{fig10}
\end{figure}

It is instructive to express $\bar{P}/s=R_sI_a^2/2$ in terms of the surface resistance $R_s$ for a stack of parallel LJJ spaced by $s$ along the $z$-axis, where $I_a=cH_a/4\pi$ is the amplitude of the ac current flowing through the LJJ. The field dependence of $R_s(H_a)=32\pi^2\bar{P}(H_a)/sc^2H_a^2$ inferred from the above results for $\bar{P}(H_a)$, is shown in  \cref{fig9}(b). Several features of $R_s(H_a)$ should be mentioned. First, $R_s(H_a)$ increases sharply above a threshold field $H_p(\omega)$ which we associate with the field onset of penetration of Josephson vortices in the junction. At small frequencies, $\alpha = \omega/\omega_c\ll 1$, the dependence $R_s(H_a)$ has a significant steplike feature component in which each step results from the change of the mean number of vortices in the LJJ by one as $H_a$ increases.  At higher frequencies, the steps $R_s(H_a)$ become less pronounced and disappear at $\alpha>1$.   For  $H_a\gg H_1$ the resistance approaches a constant value which, for an infinite LJJ, is just the surface resistance $R_0=(2\pi R_i \omega d)^{1/2}/c s$ under the normal skin effect \cite{Mc}.  However,  in our case of the LJJ of finite length ($L=20\lambda_J$), the asymptotic value of $R_s(H_a)$ is smaller than $R_0$. Results similar to those shown in  \cref{fig9}(b) were previously obtained by McDonald and Clem \cite{Mc} and by Zhai {\it et al.} \cite{Sr}.  The frequency dependence of $\bar{P}(H_a,\omega)$ at $H_a=4H_1$ shown in  \cref{fig10}, is close to the square root behavior. The latter is not surprising given that at $H=4H_1$, the Josephson vortices in the LJJ overlap and $R_s(H_a)$ shown in  \cref{fig9}(b) approaches the surface impedance of a normal conductor. 

 \subsection{Frequency dependence of penetration field}

\begin{figure}[h!]
\centering\includegraphics[width=8cm]{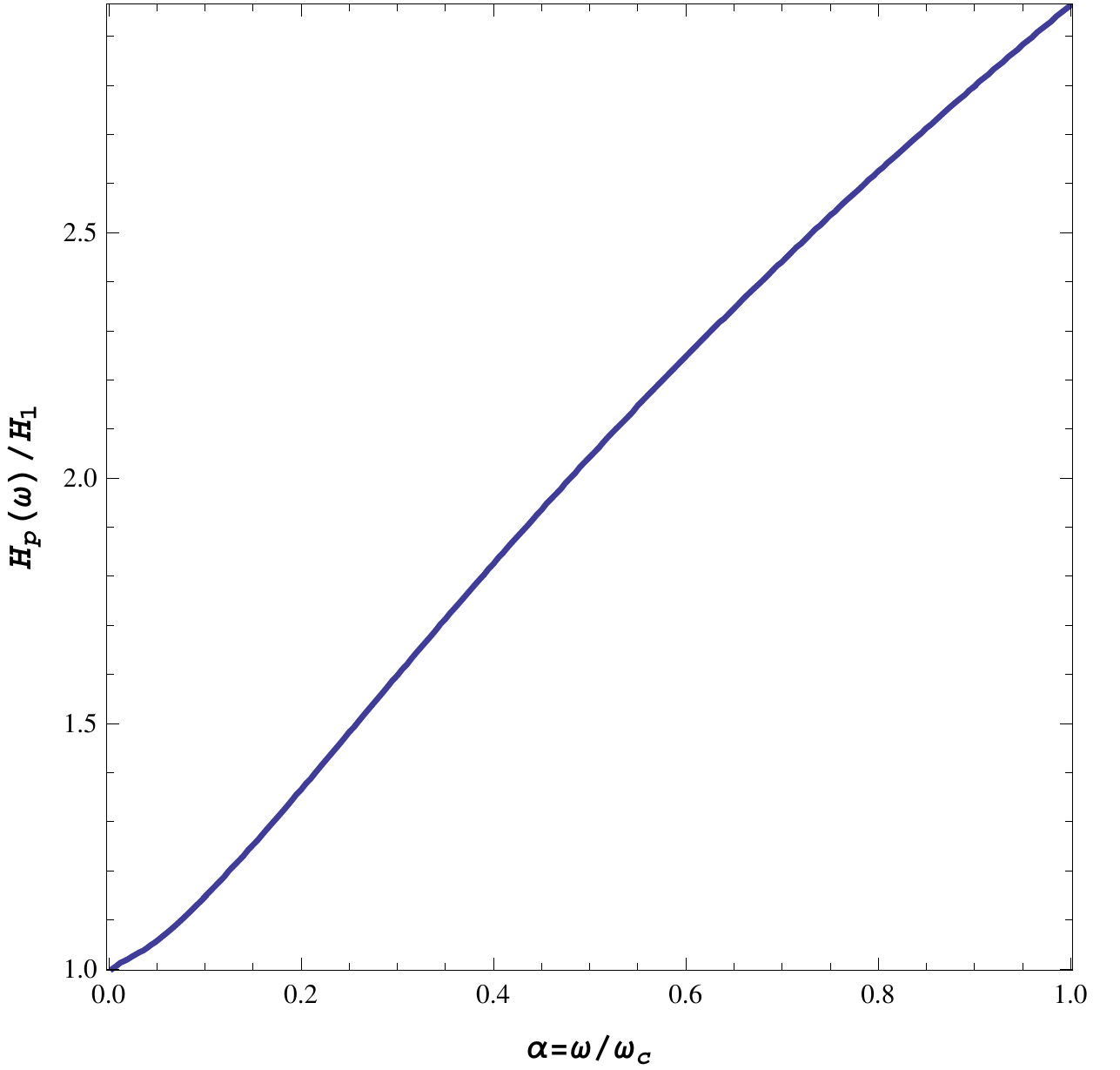}
\caption {(on the Web only) Frequency dependence of the threshold magnetic field $H_p(\omega)$ in the overdamped limit. }
\label{fig11}
\end{figure}

The field onset of sharp increase of the dissipated power $\bar{P}(H_a)$ at which the first fluxon penetrates the junction increases with the frequency of the applied field, as it is evident from  \cref{fig9}(a). The calculated frequency dependence of $H_p(\omega)$ in the overdamped regime is shown in  \cref{fig11}. Here the function $H_p(\omega)$ first increases linearly with $\omega$ at $\omega\ll\omega_c$ and then exhibits a faster increase with a downward curvature above $\alpha\sim 0.1$.  At $\omega\ll \omega_c$, the penetration field can be approximated by
 \begin{equation}
 H_p(\omega) \simeq H_1(1 +1.4\omega/\omega_c), \qquad \omega\ll \omega_c
 \label{hp}
 \end{equation} 
These results show that $H_p$ is close to the dc superheating field of the junction if $\omega\ll \omega_c$. 

\section{Asymmetric ac field}
\label{sec:asyacfield}

In the section \ref{sec:acfield} we considered a single mode ac field for which the net Lorentz force averaged over the ac period vanishes. Here we consider two situations in which 
the net force does not vanish, resulting in a preferential drift velocity of vortices. The first case is a dc magnetic field superimposed onto a single-mode ac field, and the second one is a two-mode ac field with different frequencies. In both cases the ac dynamics of vortices can be tuned by either changing the dc field $H$ or the phase shift between the two harmonics.   

\subsection{Ac driven junction biased with a dc field}

\begin{figure}[h!]
\centering\includegraphics[width=8cm]{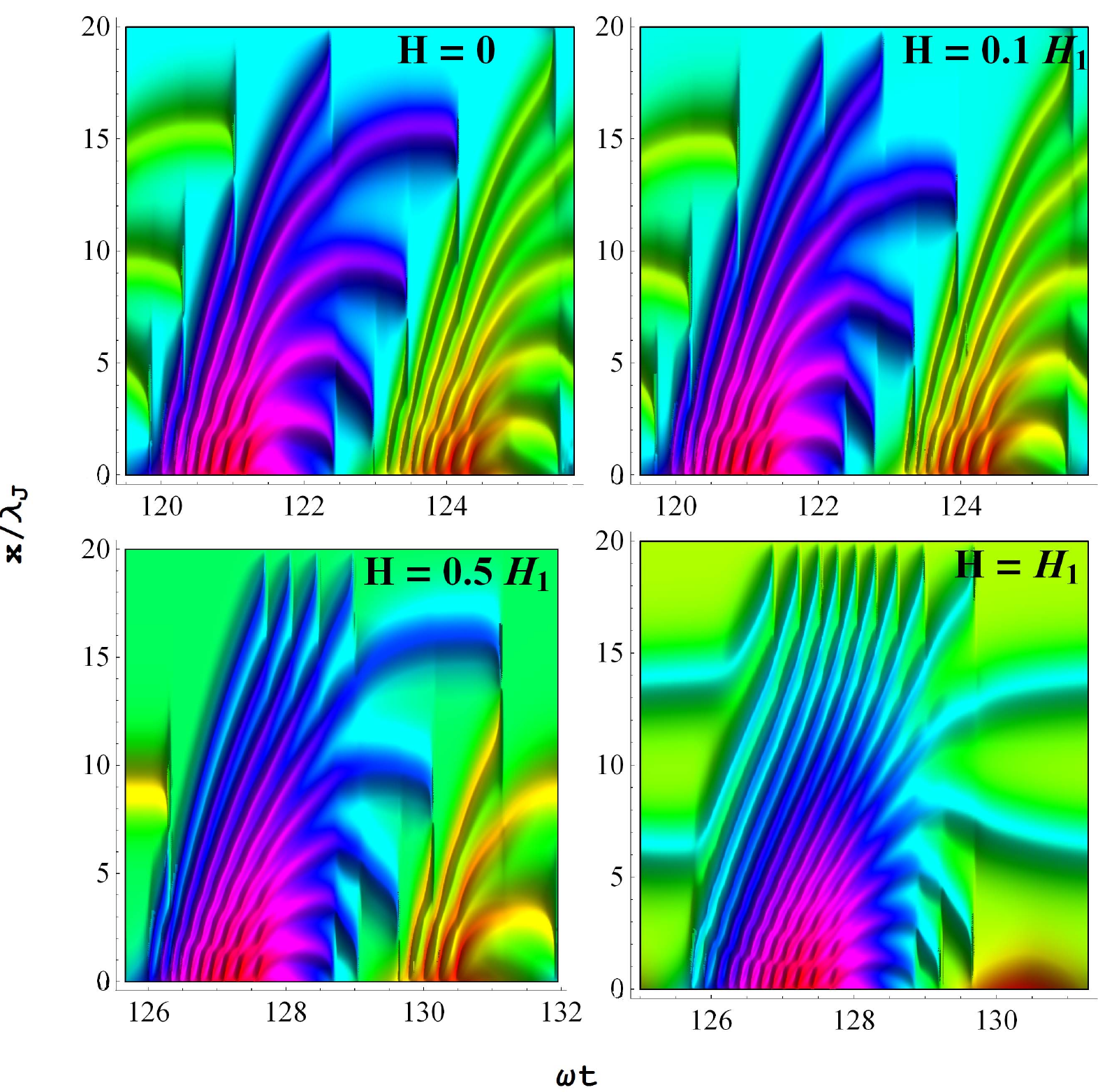} 
\caption{(on the Web only) Evolution of $B(x,t)$ along the junction during the ac period in the overdamped limit calculated for $\alpha=0.01$, $H_a=2H_1$ and different dc fields specified on the panels.}
\label{fig12}
\end{figure}

Dc magnetic field superimposed onto the ac field can result in interesting effects in LJJ which have many applications for HTS thin film junctions\cite{HLO,KS} and flux flow oscillators\cite{SS,KS}. As an illustration,  \cref{fig12} shows the results of calculations in the overdamped limit for $\alpha=0.01$, $H_a=2H_1$ and different dc field values. The main difference from the results of the section \ref{sec:acfield} is that the positive dc field breaks the symmetry between vortices and antivortices, facilitating penetration of vortices and inhibiting penetration of antivoirtices (and vice versa for negative $H$). This behavior is clearly seen in  \cref{fig12}. In the limit of $H\gg H_a$, the ac field becomes inessential, and flux dynamics approaches the unidirectional flux flow considered in Section \ref{sec:dcfield}. 

Similar to the previous sections, we define the dynamic resistance $R_s$ using the power balance $\bar{P} = R_s\langle I^2\rangle$, where the net current $I=c(H+H_a\sin t)/4\pi$  now contains both dc and ac contributions. Averaging over the ac period gives $\langle H_{tot}^2\rangle=H^2+H_a^2/2$, so that
\begin{equation}
R_s=16\pi^2 \overline{P}/c^2(H^2+H_a^2/2)
\label{rss}
\end{equation}

\begin{figure}[h!]
\centering\includegraphics[width=8cm]{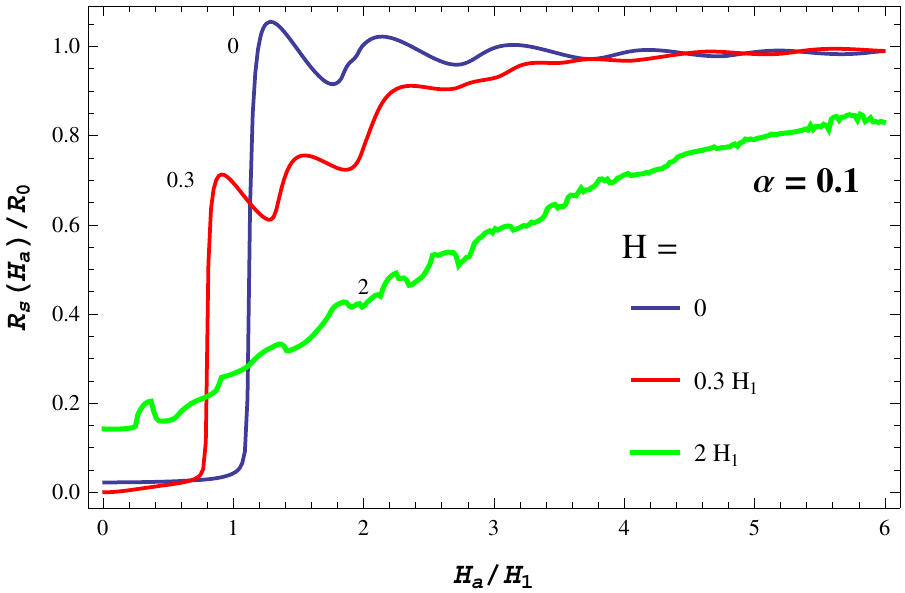} 
\caption{(on the Web only) $R_s$ dependence on $H_a$ for a long, finite junction when $\alpha=0.1$ in presence of different values of dc field. When $H_a\rightarrow H$ the ac resistance approaches the asymptotic value of $R_0$. }
\label{fig13}
\end{figure}

 Shown in  \cref{fig13} are the curves $R_s(H_a)/R_0$ calculated for $\alpha=0.1$ and different values of $H$. One can see that the dc field reduces the field threshold of vortex penetration $H_p(H)$ which is now controlled by the maximum instantaneous field value $H+H_a$. Thus, we have $H_p(H)=H_1-H$ if $H<H_1$ and $\omega\ll\omega_c$. The resistance at $H>H_1$ and $H_a\ll H$ reduces to the resistance $R$ for the unidirectional flux flow shown in  \cref{fig5}. We do not consider here a moderately dissipative case $\beta\sim\alpha$ for which the resistance $R_s$ in superimposed dc and ac fields can become negative \cite{negr}.
 
\subsection{Bi-harmonic field and the ac ratchet effect}

It is well-known that a particle driven by an external force in a periodic potential without reflection symmetry can move with a mean drift velocity $v_d$ due to the dc ratchet effect. This situation can occur in a LJJ as well if the Josephson vortex is driven by superimposed dc and ac currents \cite{negr} or by a periodic ac force containing more than one harmonics so that $v_d$ depends on the phase shift $\theta$ between two harmonics with different frequencies \cite{ratj1,ratj2,ratj3}. The dc and ac ratchet effects, and the related issues of the Brownian motors  \cite{HL,GV} have been investigated in biological systems \cite{GV}, particle separation \cite{GV,MM}, and vortex motion rectification in superconductors \cite{JS,SF1,SF2,ratdc}. 

To see how the dynamic ratchet effect can manifest itself in the LJJ geometry shown in \cref{fig1}, we consider a Josephson vortex driven by a uniform bi-harmonic current density $J(t)=J_1\cos\omega t+J_2\cos(2\omega t+\theta)$, where $\theta$ is a constant phase shift. We start with a simple model in which the vortex is treated as a particle subject to the 
ac Lorentz force, so that the velocity of the vortex $v(t)$ is described by the dynamic equation,
\begin{equation}
M\dot{v}+(1+v^2/v_0^{2})\eta v=\phi_0 J(t),
\label{eqr}
\end{equation}
where $M$ is the effective vortex mass, $\eta$ is the viscous drag coefficient \cite{KL,BP}, and the term $v^2/v_0^{2}$ describes the first nonlinear correction 
to the vortex viscosity \cite{KL,MS,LS}. We seek the solution of  \cref{eqr} in the form:
\begin{equation}
v(t)=v_d+v_1\cos(\omega t+\varphi_1)+v_2\cos(2\omega t+2\varphi_2)+v_i(t),
\label{vt}
\end{equation}
where $v_d$ is a dc drift velocity, and $v_i(t)$ is a periodic function which contains higher order harmonics. If $v(t)\ll v_0$, the nonlinear term in  \cref{eqr} is small, so that  $v_d$ can be calculated in a perturbation theory \cite{GV} by averaging  \cref{eqr} over the ac period:
\begin{equation}
v_d=-\frac{\langle v^3\rangle}{v_0^{2}}=-\frac{3v_1^{2}v_2}{4v_0^2}\cos 2(\varphi_2-\varphi_1).
\label{vdt}
\end{equation} 
Here $\langle v^3\rangle$ was calculated from  \cref{vt} neglecting the 
higher order harmonics $v_i(t)$. In the low-frequency overdamped limit $\omega\ll\eta/M$, the mass term in  \cref{eqr} can be neglected and the linearized equations $v(t)=\phi_0 J(t)/\eta$ for the first and the second harmonics yield $v_1=\phi_0 J_1/\eta$ and $v_2=\phi_0 J_2/\eta$. Then  \cref{vdt} reduces to:
\begin{equation}
v_d=-\frac{3\phi_0^{3}J_1^{2}J_2}{4v_0^{2}\eta^3}\cos\theta.
\label{vd}
\end{equation}
This relation shows that the drift velocity can be changed by varying the phase shift $\theta$ to make the vortex move either to the left or to the right. The 
case of two superimposed modes thus appears qualitatively similar to the case of superimposed dc and ac fields considered above because penetration of 
vortices can be either facilitated or inhibited by varying the phase shift $\theta$ \cite{ratj1,ratj2,ratj3}.  The case of bi-harmonic rf field can model grain boundaries in superconducting 
resonator cavities in which several resonance electromagnetic modes can be generated \cite{cav}.    

We now solve the overdamped sine-Gordon equation for a bi-harmonic field, 
$H(t)=H_{a1}\sin\omega t+H_{a2}\sin(m\omega t+\theta)$ and the boundary conditions,
\begin{equation}
\gamma_x(0,t)=h_1\sin t+h_2\sin (mt+\theta), \quad\gamma_x(l,t)=0,  
\label{bcr}  
\end{equation}
where $\left\lbrace  h_1,h_2 \right\rbrace  =(4\pi\lambda\lambda_J/\phi_0)\left\lbrace  H_{a1},H_{a2} \right\rbrace$ and $m$ is integer.

\begin{figure}[h!]
\centering\includegraphics[width=8cm]{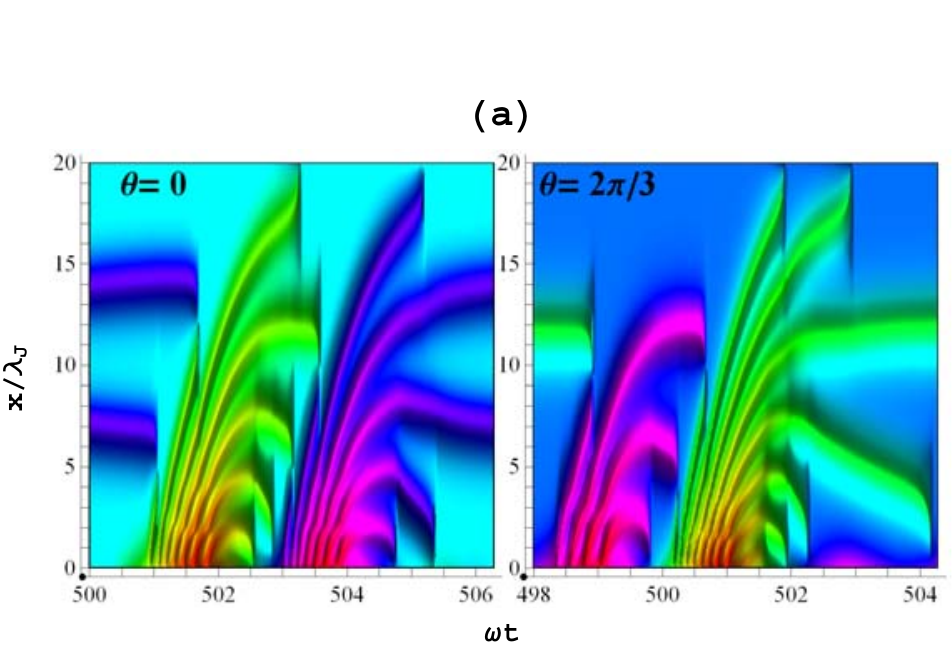} 
\bigskip
\centering\includegraphics[width=8cm]{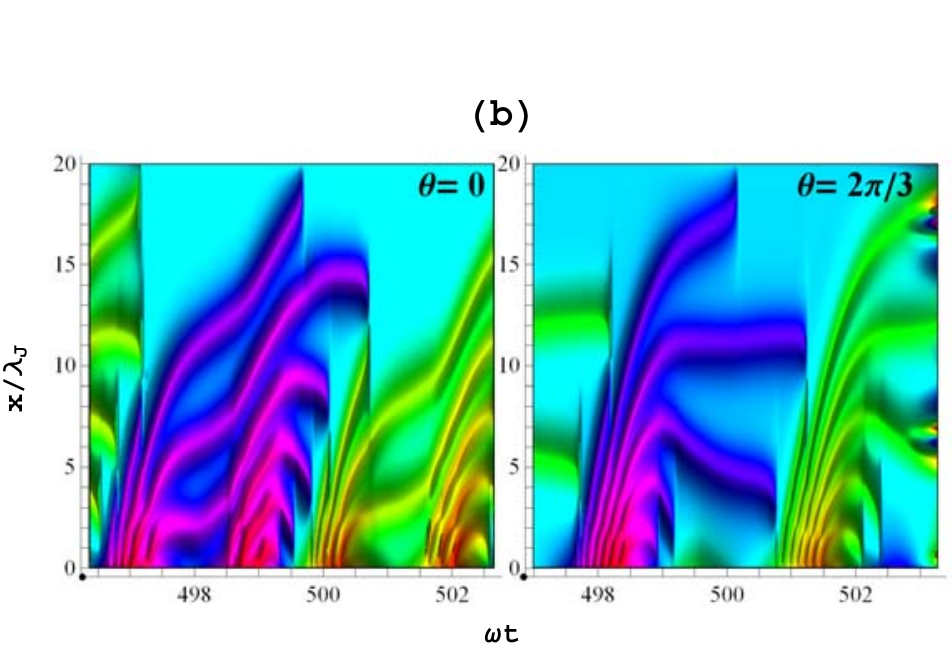}
\caption{(on the Web only) Evolution of $B(x,t)$ for a $2\pi$ period of the bi-harmonic magnetic field.  (a): $m=2$, $H_{a2}/H_{a1}=0.66$ and $\alpha=0.01$. Changing $\theta$ from 0 to $2\pi/3$ decreases the number of vortices and increases the number of anti-vortices in the junction. (b): $m=3$, $H_{a2}/H_{a1}=0.66$ and $\alpha=0.01$. Varying $\theta$  affects flux dynamics but does not result in the vortex/antivortex imbalance. }
\label{fig14}
\end{figure}

In the overdamped limit the solutions for $\gamma(x,t)$ have the same periodicity as $H(t)$. For even $m$, the ratio of the mean numbers of fluxons and antifluxons can be tuned by varying $\theta$, which was observed in Ref. \cite{SF1,SF2}.  For instance, as  \cref{fig14}(a) shows, increasing $\theta$ from 0 to $2\pi/3$ inhibits penetration of vortices and facilitates penetration of anti-vortices into the junction. For odd values of $m$, the field satisfies the condition $H(t+T/2)=-H(t)$ so the change of $\theta$ does not result in the vortex/antivortex imbalance, although flux dynamics is affected by $\theta$.  As an example,  \cref{fig14}(b) shows that varying $\theta$ from 0 to $2\pi/3$ affects the dynamics of $B(x,t)$ symmetrically for both fluxons and anti-fluxons. 

\begin{figure}[h!]
\centering\includegraphics[width=7.5cm]{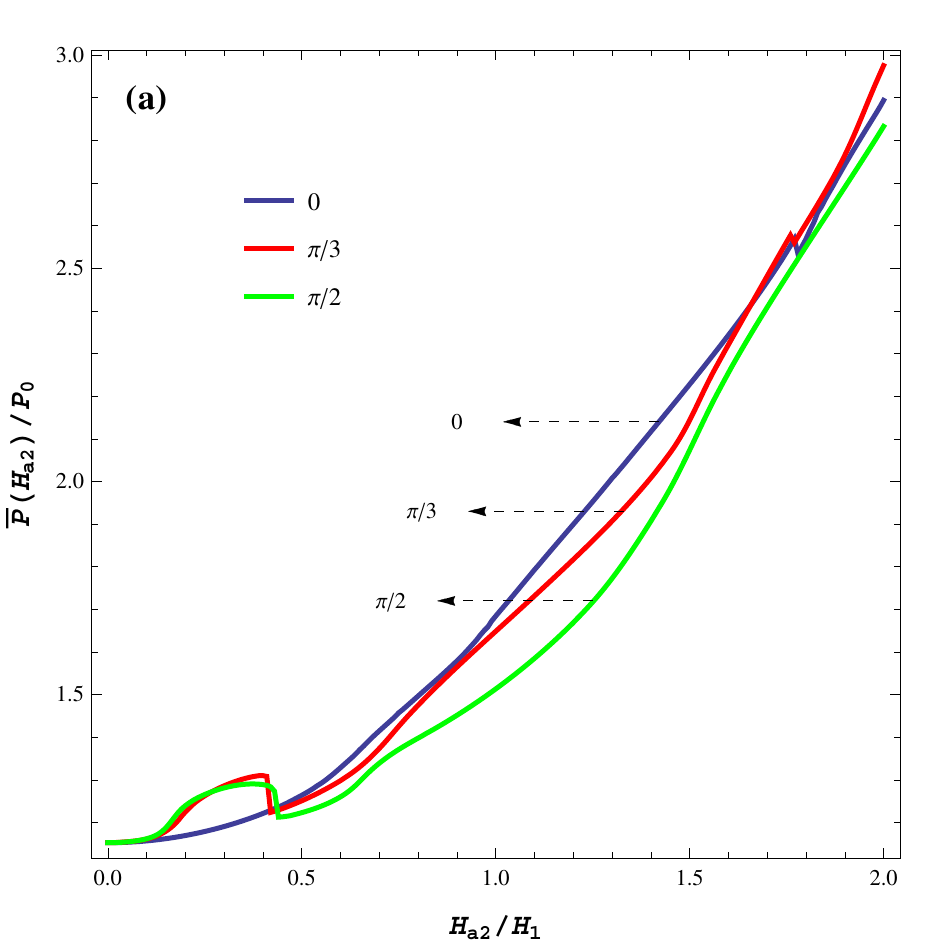}
\bigskip
\centering\includegraphics[width=7.5cm]{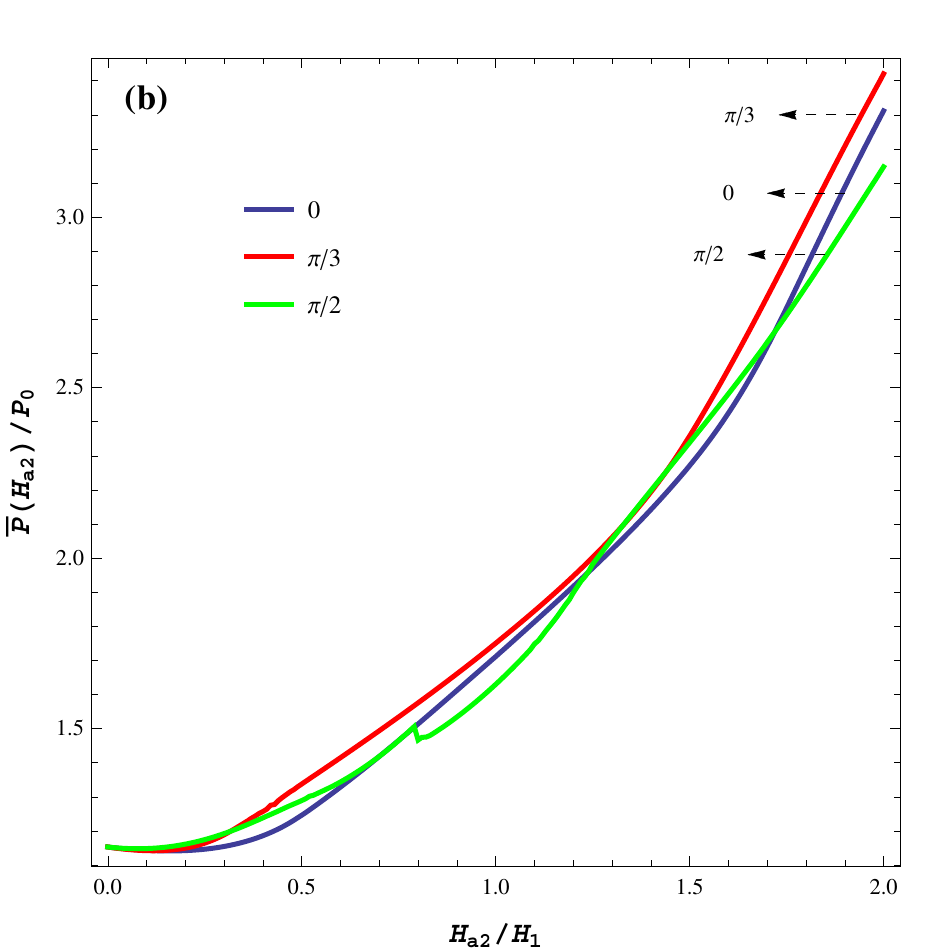}
\caption{(on the Web only) Plots of the average dissipated power $\overline P$ as a function of second mode amplitude $H_{a2}$, for different values of relative phase $\theta$ while $\alpha=0.05$ and the fundamental mode is kept constant at $H_{a1}=2H_1$. (a) $m=2$. (b) $m=3$.}
\label{fig15}
\end{figure}

 \cref{fig15} shows how the dissipated power can be tuned by varying $H_{a2}$ and $\theta$ for the fixed amplitude of the first harmonic, $H_{a1}$. For $m=2$, the power $\bar{P}$ generally increases with $H_{a2}$ but the change of $\theta$ from $0$ to $\pi/2$ results in humps and dips on the curve of $\bar{P}(H_{a2})$ at $H_{a2}\approx H_1/2$ and $H_{a2}\approx 1.8H_1$. For $m=3$, the dips in $\bar{P}(H_{a2})$ are less pronounced but still apparent at $H_{a2}\approx 0.8H_1$ and $\theta=\pi/2$. In both cases changing $\theta$ from $0$ to $\pi/2$ can reduce $\bar{P}(H_{a2})$ in certain regions of $H_{a2}$ while increasing $\bar{P}(H_{a2})$ in others.

\begin{figure}[h!]
\centering\includegraphics[width=7.5cm]{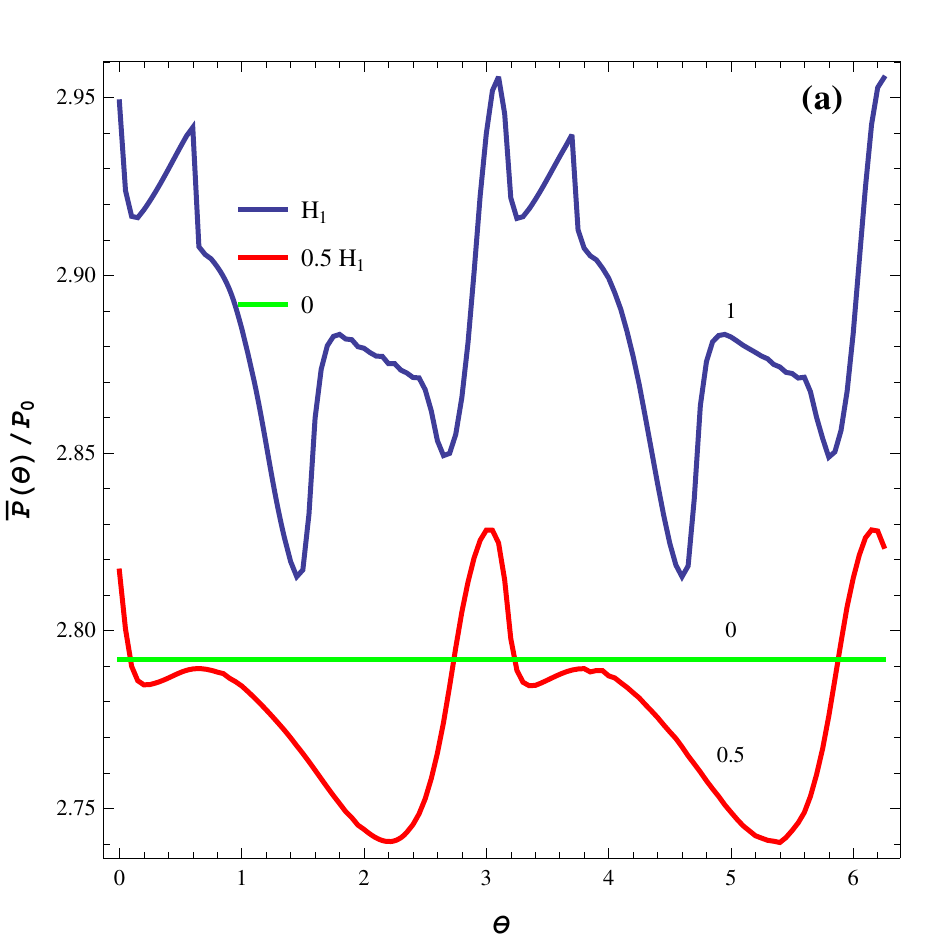} 
\bigskip
\centering\includegraphics[width=7.5cm]{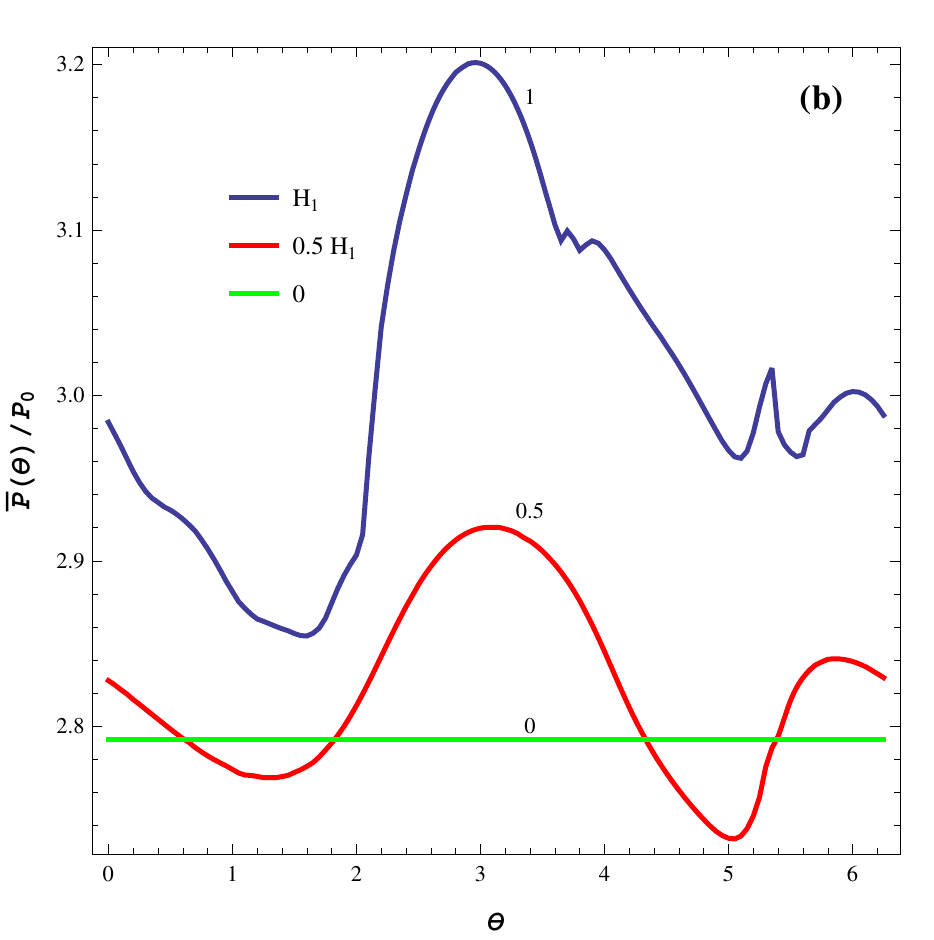}
\caption{(on the Web only) Plots of the average power $\overline P$, as a function of the relative phase $\theta$, for $\alpha=0.05$, fixed total field magnitude $(H_{a1}^2 + H_{a2}^2)^{1/2}=3H_1$ and different values of $H_{a2}$:  (a) $m=2$ and (b) $m=3$ . The solid horizontal line at $\approx 2.79$ in both plots shows $\bar{P}$ 
at $H_{a1}=3H_1$ and $H_{a2}=0$.}
\label{fig16}
\end{figure}

The dependencies of $\bar{P}(\theta)$ for $m=2$ and $m=3$, at a fixed field amplitude $(H_{a1}^2+H_{a2}^2)^{1/2}=3H_1$ and different values of $H_{a2}$ are shown in  \cref{fig16}(a) and (b). Here $\bar{P}(\theta)$ can vary rapidly with $\theta$, although the maximum change of $\bar{P}(\theta)$ does not exceed $10\%$. Interestingly, the most pronounced reduction of $\bar{P}(\theta)$ in the suitable ranges of $\theta$ occurs if the amplitude of the second harmonics is small as compared to $H_{a1}$. 
 
\section{Discussion}
\label{sec:diss}

The results of this work show that the electromagnetic response of a long but finite Josephson junctions in ac magnetic fields 
can be quite complicated due to penetration, oscillation and annihilation of Josephson vortices. The nonlinear dynamics of Josephson vortices results in essential dependencies of the averaged dissipated power $\bar{P}$ and the dynamic resistance $R$ on the field amplitude. Here  $\bar{P}(H)$ and $R(H)$ can have steps and peaks due to the change of the number of trapped vortices in the junction as $H$ increases. The calculated field dependence of  the surface resistance $R(H)$ is far from linear, inconsistent with the model  assumptions of previous works \cite{gbi4,gbi5}. It is important to point out that $\bar{P}(H)$ is obtained by averaging the instant power $P(t,H)$ over the ac period during which $P(t)$ has strong spikes due to annihilation of vortices and antivortices in the junction, the magnitude of these power spikes can be much higher than the smooth background contribution to $P(t)$.  In high-$J_c$ Josephson junctions these power spikes may trigger thermo-magnetic instabilities in the rf field \cite{ap}.

The penetration of Josephson vortices occurs above the threshold field $H_p(\omega)$ at which the dissipated power increases significantly. If the local $J_c$ at the edge of the junction is not reduced by materials defects, the dynamic penetration field $H_p(\omega)$ remains close to the dc Josephson superheating field of the Meissner state $H_1=\phi_0/2\pi\lambda\lambda_J$ if $\omega/\omega_c \ll 1$. The characteristic frequency $\omega_c $ in  \cref{om} is proportional to the product $J_cR_i$, so the frequency dependence of $H_p(\omega)$ is most pronounced for low-$J_c$ and low resistance junctions. The estimates given above show that for the grain boundaries in Nb, the frequency-dependent correction in $H_p(\omega)$ is small for 
$\omega < \Delta/\hbar$, where $\Delta$ is the superconducting gap. 

As was mentioned in the introduction, a LJJ of finite length can model the electromagnetic response of grain boundaries in polycrystalline superconductors.  The grain boundaries in Nb resonator cavities appear to be strongly coupled and do not behave as the conventional Josephson junctions up to very high magnetic fields at which densities of screening current become of the order of the depairing current density. As a result, the field onset of penetration of mixed Abrikosov-Josephson vortices \cite{nje1,nje2,nje3} is close to the lower critical field of intra-grain vortices $H_{c1}\simeq 170$ mT.  By contrast, the grain boundaries in Nb$_3$Sn, iron-based superconductors or high-$T_c$ cuprates do behave as Josephson weak links \cite{GB,DG}, so one can expect that the nonlinear effects addressed in this paper can manifest themselves in the surface impedance at rather low fields $H\simeq H_1<H_{c1}$. Such effects can also be essential for the rf performance of polycrystalline multilayer screens which were suggested to enhance the breakdown field of Nb resonator cavities \cite{ml}.  

\section*{Acknowledgments}
This work was supported by the US Department of Energy, Division of High Energy Physics under grant No. DE-SC0010081.

\section*{References}
\bibliographystyle{elsarticle-num}
\bibliography{mybib}

\end{document}